\documentclass{iopart}

\usepackage{iopams,amsfonts,amssymb,amsthm} 
\usepackage{geometry}                
\usepackage{graphicx}
\usepackage{rawfonts}
\usepackage{amssymb}
\usepackage{epstopdf}
\usepackage[usenames,dvipsnames,svgnames]{xcolor}
\usepackage{color}
\usepackage[colorlinks=true,citecolor=blue,linkcolor=red,urlcolor=blue]{hyperref}
\usepackage{multirow}
\usepackage{multicol}
\usepackage{comment}

\input prepictex.tex
\input pictex.tex
\input postpictex.tex

\newtheorem{theo}{Theorem}
\newtheorem{lemm}{Lemma}

\def\BRef#1{{(}\ref{#1}{)}}

\def\qed{$\Box$}

\def\NatN{\mathbb{N}}

\def\exp#1{\hbox{exp($#1$)}}

\begin{document}

\title[Combs]{Exponential growth rate of lattice comb polymers}

\author{E J Janse van Rensburg$^\star$ \& S G Whittington$\dagger$}

\address{
$^\star$ Department of Mathematics and Statistics, York University, Toronto,  M3J 1P3, Canada\\ 
$\dagger$ Department of Chemistry,  University of Toronto, Toronto,  M5S 3H6, Canada}
\ead{
$^\star$rensburg@yorku.ca,
$^\dagger$stuart.whittington@utoronto.ca}
\vspace{10pt}
\begin{indented}
\item[] \today
\end{indented}

\begin{abstract}
{We investigate a lattice model of comb polymers and derive bounds on the exponential growth rate of the number of embeddings of
the comb.  A comb is composed of a backbone that is a self-avoiding walk and a set of $t$ teeth, also modelled as mutually and self-avoiding 
walks, attached to the backbone at vertices or nodes of degree 3.  Each tooth of the comb has $m_a$ edges and there are $m_b$ edges in the backbone between adjacent 
degree 3 vertices and between the first and last nodes of degree 3 and the end vertices of degree 1 of the backbone.  We are interested in the exponential growth rate as 
$t \to \infty$ with $m_a$ and $m_b$ fixed.  We prove upper bounds on this growth rate and show that for small values of $m_a$ the growth rate is strictly
less than that of self-avoiding walks.}

\end{abstract}


\vspace{2pc}

%
\maketitle
%
%

\section{Introduction}
\label{sec:Introduction}

Comb polymers are a class of branched polymers with simple connectivity and yet useful and
interesting properties.  Generally, a comb is a long (linear) backbone polymer with side chains
(called teeth) substituted at regular intervals along the backbone.  Copolymeric combs are useful
as steric stabilizers of dispersions \cite{Xie2017}.  Usually the backbone of the comb adsorbs
on a colloidal particle and the teeth extend in solution.  These teeth lose conformational 
degrees of freedom when another particle approaches, resulting in a repulsive entropic force
between particles.  

\begin{figure}[h!]
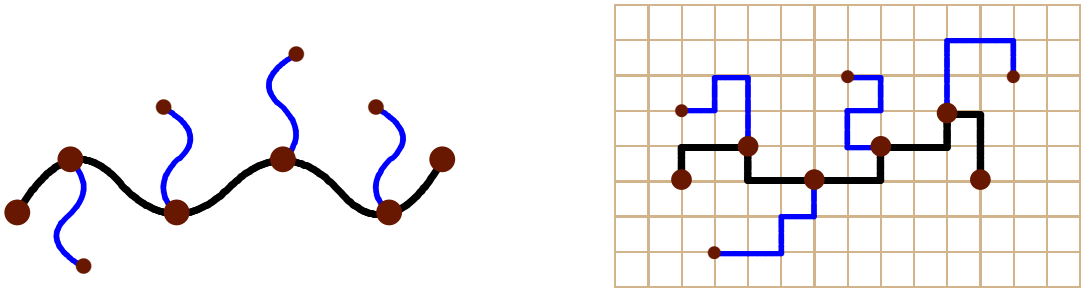

    \beginpicture
\setcoordinatesystem units <1pt,1pt>
\setplotarea x from -10 to 170, y from -30 to 30

\color{Tan}
\setplotsymbol ({\scalebox{1.0}{$\bullet$}})

\setplotsymbol ({\scalebox{0.67}{$\bullet$}})
\setquadratic
\color{black}
\plot 0 -10 20 10 40 0 60 -10 80 0 100 10 120 0 140 -10 160 10  /
\setplotsymbol ({\scalebox{0.50}{$\bullet$}})
\color{blue}
\plot 20 10 25 0 20 -10 15 -20 25 -30 /
\plot 60 -10 55 0 60 10 65 20 55 30 /
\plot 100 10 105 20 100 30 95 40 105 50 /
\plot 140 -10 135 0 140 10 145 20 135 30 /
\setlinear

\color{Sepia}
\multiput {\scalebox{2.5}{$\bullet$}} at 0 -10 160 10 20 10 60 -10 100 10 140 -10 /
\multiput {\scalebox{1.5}{$\bullet$}} at 25 -30 55 30 105 50 135 30  /

\setcoordinatesystem units <1.25pt,1.25pt> point at -200 0 
\setplotarea x from -20 to 120, y from -30 to 55
\color{Tan}
\setplotsymbol ({\scalebox{0.2}{$\bullet$}})
\grid 14 8

\setcoordinatesystem units <1.25pt,1.25pt> point at -200 -2 
\color{black}
\setplotsymbol ({\scalebox{0.67}{$\bullet$}})
\plot  0 0 0 10 20 10 20 0 60 0 60 10 80 10 80 20 90 20 90 0  /
\setplotsymbol ({\scalebox{0.50}{$\bullet$}})
\color{blue}
\plot 20 10 20 31 10 31 10 21 0 21  /
\plot 40 0 40 -11 30 -11 30 -22 20 -22 10 -22 /
\plot 60 10 50 10 50 21 60 21 60 31 50 31 /
\plot 80 20 80 42 100 42 100 31 /

\color{Sepia}
\multiput {\scalebox{2.0}{$\bullet$}} at 0 0 20 10 40 0 60 10 80 20 90 0  /
\multiput {\scalebox{1.25}{$\bullet$}} at 0 21 10 -22 50 31 100 31  /

\color{black}\normalcolor
\endpicture

\caption{(Left) A schematic of a comb polymer consisting of a linear polymer 
backbone with linear polymer teeth grafted at regular intervals along the
backbone.  (Right) A square lattice model of a comb.  The conformations
of the lattice comb model the conformational degrees of freedom of the
comb polymer.  In this case, $t=4$, $m_a=5$ and $m_b=3$.}
\label{F1}
\end{figure}

In this paper we consider self-avoiding walk models of square and cubic lattices combs.  In
figure \ref{F1} a schematic drawing of a comb, and its representation in the square lattice,
are shown.  Teeth in the lattice comb are attached at \textit{nodes} to the backbone, and
these nodes are uniformly spaced along the backbone.  The length of each tooth is $m_a$,
while the nodes are spaced a distance $m_b$ apart along the backbone.  If there are $t$
teeth substituted along the backbone, then we say the \textit{degree} of the comb is $t$,
and the backbone has length $(t{+}1)\,m_b$ while the size of the comb is $(t{+}1)\,m_b+t\,m_a$.
This lattice model is of a comb in a dilute solution in a good solvent, and the 
conformational degrees of freedom in the lattice quantify its conformational entropy. 
For a general review of the properties of combs and related branched polymers see 
\cite{Potemkin}.

Much of the theoretical work on lattice combs has focussed on their dimensions in a dilute 
solution (for example, see references \cite{Lipson1991,Lipson1993,Potemkin} and other references 
therein).  A central question about lattice combs is the exponential rate of increase of the
number of conformations (or embeddings) of the comb as a function of its total size.  This
question was considered in various ways in references \cite{Gaunt,RensburgSoterosWhittington,
Rensburg2019,Lipson1993}.  We shall be concerned about this question as well,
but in the context explored in references \cite{RensburgSoterosWhittington} 
(see especially section 5 therein) and \cite{Lipson1993}. 

If the number of teeth is fixed at $t=2$ and $m_a=m_b=m \to \infty $ then \cite{Gaunt} 
the exponential growth rate is the same as that of self-avoiding walks.  In 1993
Lipson \cite{Lipson1993} used Monte Carlo methods and combinatorial upper bounds 
to argue that, if $m_a$ and $m_b$ are fixed but $t$ goes to infinity, then 
the exponential growth rate is strictly less than that of self-avoiding walks.  The 
main result in this paper is a proof of this when the teeth are very short.  
We shall use and extend some of the ideas that appeared 
in \cite{Lipson1993}, and also supplement our results with numerical data.

We denote the number of self-avoiding walks of length $n$ steps from the origin 
in a lattice by $c_n$, and recall that the \textit{connective constant} of 
self-avoiding walks is defined by \cite{Hammersley1957}
\begin{equation}
\kappa = \lim_{n\to\infty} \frac{1}{n} \log c_n = \inf_{n>0} \frac{1}{n}  \log c_n.
\label{eqn:sawlimit}
\end{equation}
The \textit{growth constant} $\mu$ of self-avoiding walks is defined by 
$\mu = e^\kappa$. 

The connective constant is lattice dependent, and 
we shall sometimes add a subscript to denote the lattice, for example, 
$\kappa_d$ for the connective constant of the $d$-dimensional hypercubic 
lattice (that is, $\kappa_2$ for the square lattice, and $\kappa_3$ for the
cubic lattice), or $\kappa_H$ for the hexagonal (honeycomb) lattice.  We use similar
subscripts for $\mu$.

For most lattices the values of $\kappa$ or $\mu$ are not known exactly.  Best
estimates for $\mu$ have been determined by extrapolating from exact counts
of self-avoiding walks.  For example
\begin{equation}
\mu =
\cases{
2.6381585321(4),& \hbox{square lattice \cite{CJ12,Jensen13}}; \cr
4.150797226(26),& \hbox{triangular lattice \cite{Jensen2004}}; \cr
2.560576765(10),& \hbox{Kagom\'e lattice (attributed to I Jensen - \cite{Jensen2004})}; \cr
4.684039931(27),& \hbox{simple cubic lattice \cite{C13}} . \cr
}
\label{exact:mu}
\end{equation}
In the case of the hexagonal lattice $\mu$ is known exactly, first determined
in reference \cite{N82}, and then proven in \cite{Smirnov}: 
\begin{equation}
\mu_H = \sqrt{2+\sqrt{2}} = 1.847759 \ldots
\label{eqn:hexagonalsaw}
\end{equation}
This result will be useful in section \ref{sec:hexagonal}.  For other lattices rigorous 
lower and upper bounds are available \cite{ConwayGuttmann,FisherSykes,Jensen2004},  
and we shall make use of some of these bounds.  The lower bounds useful to
us in this paper are
\begin{equation}
\mu >
\cases{
2.625622,& \hbox{square lattice \cite{Jensen2004}}; \cr
4.118935,& \hbox{triangular lattice \cite{Jensen2004}}; \cr
2.548497,& \hbox{Kagom\'e lattice \cite{Jensen2004}}; \cr
4.225,& \hbox{simple cubic lattice \cite{FisherSykes}} .
}
\label{lower:mu}
\end{equation}
In this paper we shall be concerned with the connective constant of lattice combs,
and in particular compare bounds on the connective constants of combs to the 
connective constants of self-avoiding walks in the same lattice.

In a lattice with coordination number $\gamma$ one can obtain upper bounds 
on $\mu$ by noting that $c_{nm} \leq (c_n)^m((\gamma{-}1)/\gamma)^{m-1}$ 
(concatenate $m$ walks of length $n$ and note that each time a walk is added, 
then there are $\gamma{-}1$ directions available for its first step).  Taking logarithms 
of this, and dividing by $nm$, and then $m\to\infty$ with $n$ fixed,
\begin{equation}
\log \mu \leq (\log c_n + \log((\gamma{-}1)/\gamma))/n .
\end{equation}
In the square lattice and for $n=1$ one obtains $\mu \leq 3$, and the bound
improves as $n$ increases.  For $n=79$ \cite{CJ12,Jensen13} it gives $\mu < 2.6848$.
In the cubic lattice, $n=36$ \cite{SBB11} gives the bound $\mu < 4.7587$.  We
shall use generalizations of this approach to find numerical upper bounds on the limiting
free energy of combs and brushes. 

We write $g(m_a,m_b,t)$ for the number of combs (\emph{ie} 
the number of lattice embeddings) with $t$ teeth, each of length $m_a$, with 
$m_b$ edges between adjacent pairs of degree 3 vertices along the backbone 
and with $m_b$ edges between the first and last degree 3 vertices and the 
end vertices of degree $1$ of the backbone.  

In \cite{RensburgSoterosWhittington} it was shown that the limit
\begin{equation}
\zeta(m_a,m_b) = \lim_{t \to \infty} \frac{1}{(m_a{+}m_b)\,t} \log g(m_a,m_b,t)
\label{eqn:comblimit}
\end{equation}
exists.  The proof was for the simple cubic lattice but the same kind of concatenation 
argument works for other lattices.  The function $\zeta(m_a,m_b)$ is the free energy 
(or \textit{connective constant}) of an infinite comb with backbone segments all 
of length $m_b$, and teeth all of length $m_a$. If $m_a=0$ and $m_b>0$ then 
clearly, $\zeta(0,m_b) = \log \mu=\kappa$.  Generally the relations between 
the $\zeta(m_a,m_b)$ for various finite values of $m_a$ and $m_b$ are not known.  
For example,  one may expect that $\zeta(1,1) \leq \zeta(2,1)$, but a proof of 
this is not known.  However, there are cases when $\zeta(m_a,m_b) < \kappa$, 
and these cases are our concern.   In this paper we show this for various values 
of $m_a$ and $m_b$.

The outline of the paper is as follows.  In section \ref{sec:background} 
we review some results about self-avoiding walks and lattice stars that 
we shall need later.  The remainder of the paper is concerned with 
the exponential growth rate of the number of combs as $t \to \infty$.  
We begin in section \ref{sec:long} where 
we make use of a pattern theorem for self-avoiding walks \cite{Kesten} 
to show that the growth constant of combs is strictly less than that of 
self-avoiding walks when the backbone segments are long enough, but finite, 
and the teeth are very short. 
In sections \ref{sec:growthrate} and \ref{sec:stars} we derive upper 
bounds on the growth rate of the number of combs when the teeth 
are short, and show that these establish in many cases that the growth 
rate is strictly less than that of self-avoiding walks.  In section 
\ref{sec:brushes} we extend some of these approaches to brushes, 
where there is more than one tooth at each branch point, while in 
section \ref{sec:limiting} we consider the situation where $m_a$ or $m_b$ 
goes to infinity, after the limit $t \to \infty$ is taken.

\section{Some background results and notation}
\label{sec:background}

\subsection{Self-avoiding walks and lattice combs}

By equation (\ref{eqn:sawlimit}) $c_n \geq \mu^n$ generally so that the 
sequence $(c_n^{1/n})$ approaches $\mu$ from above (that is, 
$\frac{1}{n} \log c_n \geq \kappa$).  While not much is known about 
the rate of approach to the limit in (\ref{eqn:sawlimit}),  Hammersley and 
Welsh \cite{HammersleyWelsh} showed for the $d$-dimensional hypercubic 
lattice that for any $\gamma_0 > \pi \sqrt{2/3}$ there exists $n_0(\gamma_0)$ such that 
\begin{equation}
\mu^n \le c_n \le A \mu^n e^{\gamma_0 \sqrt{n}},
\qquad\hbox{for all $n>n_0(\gamma_0)$}.
\label{eqn:Hammersley-Welsh}
\end{equation}
The constant $A$ is lattice dependent.

An important result about self-avoiding walks that we shall use in 
section \ref{sec:long} is Kesten's pattern theorem \cite{Kesten}. 
A \emph{pattern} is any fixed self-avoiding walk.  If there exists a self-avoiding 
walk on which the pattern occurs three times \cite{HammersleyWhittington} 
then the pattern is a \emph{Kesten pattern}.  Suppose that $P$ is a Kesten 
pattern and suppose that $c_n(\overline{P})$ is the number of $n$-edge 
self-avoiding walks on which the Kesten pattern $P$ never occurs.  Then \cite{Kesten}
\begin{equation}
\lim_{n\to \infty} \frac{1}{n} \log c_n(\overline{P}) 
  = \kappa(\overline{P}) < \kappa.
\label{eqn:pattern}
\end{equation}
That is, a Kesten pattern appears at least once on all except exponentially 
few sufficiently long self-avoiding walks. Additional results about self-avoiding 
walks can be found in references \cite{Rensburg2015,MadrasSlade}.  

A \emph{lattice $f$-star} or \emph{f-star} is a model of a star polymer with 
$f$ arms, and it is an embedding of a tree with one vertex of degree $f$ (called 
the \emph{central node} of the star) and $f$ vertices of degree $1$, in a lattice 
with coordination number at least $f$.
We denote the length of the $i$-th arm of the star by $n_i$, 
so that $n=n_1+n_2+\cdots+n_f$ is the total length or size of the star.  If $n_i=n_j$ 
for all $1\leq i,j,\leq f$, then the star is \textit{uniform}.  In this case the total length 
of the star is $n=mf$, for some $m\in\NatN$.  If $n$ is not a multiple of $f$, 
then a star can be \emph{almost uniform} if $\lfloor n/f \rfloor \leq n_i \leq \lceil n/f \rceil$.  
We shall write $s_n(f)$ for the number of embeddings of a uniform or almost
uniform $f$-star with total length $n$, and we note the important fact that the
star's central node is rooted at the origin.  Then \cite{Wilkinson}
\begin{equation}
\lim_{n\to\infty} \frac{1}{n} \log s_n(f) = \kappa.
\end{equation}
More generally, stars may be non-uniform and we denote the number of $f$-stars 
with arms of lengths $(n_1,n_2,..,n_f)$ by $s(n_1,n_2,\ldots,n_f)$.  In the case that
$n_i=n_j=m$ then $s(m,m,\ldots,m)=s_{mf}(f)$ is the number of uniform $f$-stars 
of size $n=mf$.

In the case of simple cubic lattice combs it can be shown that 
\begin{eqnarray}
\label{eqn:starlimit}
\lim_{{m_a,m_b \to\infty}\atop{m_b=o(m_a)}}
 \frac{1}{tm_a+ (t{+}1)m_b} \log g(m_a,m_b,t)=& \kappa_3  \\
\label{eqn:walklimit}
\lim_{{m_a,m_b\to\infty}\atop{m_a=o(m_b)}} 
\frac{1}{tm_a+(t{+}1)m_b} \log g(m_a,m_b,t)  =& \kappa_3.
\end{eqnarray}
Lower bounds follow from theorems 2 and 3 in \cite{RensburgSoterosWhittington}
setting $a=b=y=1$ and using the fact that combs confined to a half-space and 
terminally attached to the dividing plane are a subset of all combs.  Upper 
bounds follow by regarding the teeth as independent of one another and 
independent of the backbone, giving the inequality
\begin{equation}
g(m_a,m_b,t) \le c_{m_b(t+1)}\, c_{m_a}^t.
\end{equation}
Taking logarithms, dividing by $tm_a+(t{+}1)m_b$, and taking the 
appropriate limits with $t$ fixed, gives the required upper bounds.    A similar 
argument, using theorem 1 in \cite{RensburgSoterosWhittington} with 
$a=b=y=1$ shows that
\begin{equation}
\label{eqn:uniformlimit}
\lim_{m\to\infty}  \frac{1}{(2t{+}1)m} \log g(m,m,t) = \kappa_3.
\end{equation}
The expressions in equations (\ref{eqn:starlimit}),  (\ref{eqn:walklimit}) and
(\ref{eqn:uniformlimit}) are independent of $t$ and the limit $t \to \infty$ 
can be taken  \emph{after} the limits $m_a,m_b \to \infty$ have been 
taken, with the same result. 

In reference \cite{RensburgSoterosWhittington} equations (\ref{eqn:starlimit}) 
and (\ref{eqn:walklimit}) correspond to the \emph{star} and 
\emph{self-avoiding walk} limits of lattice combs.  The limit as $t\to\infty$ 
shown in equation (\ref{eqn:comblimit}), defining $\zeta(m_a,m_b)$,  is 
the limit as the number of teeth, $t$, increases with $m_a$ and $m_b$ fixed. 
In the next sections we consider the situation when the limit $t\to\infty$ is 
taken first, leaving $m_a$ or $m_b$ finite.  We shall see that in these cases 
the limits are in general functions of $m_a$ and $m_b$, and in some cases 
are strictly less that $\kappa_2$ in the square lattice, and strictly less than 
$\kappa_3$ in the cubic lattice (this may also be the case in other lattices).

\subsection{Notation and branched structures}

\begin{figure}[h!]
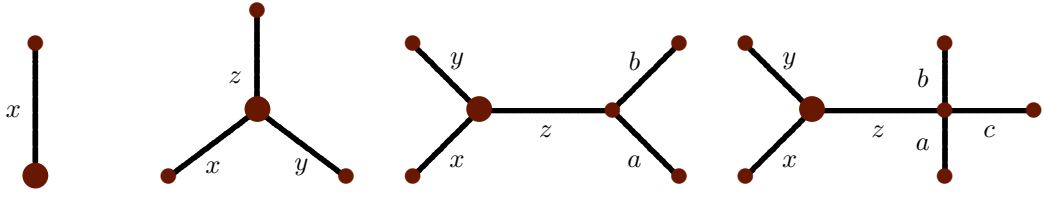

\beginpicture
\setcoordinatesystem units <1.67pt,1.25pt>
\setplotarea x from 0 to 270, y from -10 to 30

\setplotsymbol ({\scalebox{0.50}{$\bullet$}})
\color{black}
\plot 20 0 20 40  /
\put {$x$} at 15 20    
\color{Sepia}
\multiput {\scalebox{1.5}{$\bullet$}} at 20 40    /
\multiput {\scalebox{2.5}{$\bullet$}} at 20 0   /

\setcoordinatesystem units <1.67pt,1.25pt> point at -50 0 
\setplotsymbol ({\scalebox{0.50}{$\bullet$}})
\color{black}
\plot  0 0 20 20 40 0 /
\plot  20 20 20 50 /
\put {$x$} at 10 3   \put {$y$} at 30 3    \put {$z$} at 15 30    
\color{Sepia}
\multiput {\scalebox{1.5}{$\bullet$}} at 0 0 40 0 20 50   / 
\multiput {\scalebox{2.5}{$\bullet$}} at 20 20   / 

\setcoordinatesystem units <1.67pt,1.25pt> point at -100 0
\setplotsymbol ({\scalebox{0.50}{$\bullet$}})
\color{black}
\plot 5 0 20 20 5 40  /
\plot 20 20 50 20  /
\plot 65 40 50 20 65 0 /
\put {$z$} at 35 14   \put {$x$} at 15 5    \put {$y$} at 15 35    
\put {$a$} at 55 5  \put {$b$} at 55 35
\color{Sepia}
\multiput {\scalebox{1.5}{$\bullet$}} at 5 0 5 40 50 20 65 0 65 40  /
\multiput {\scalebox{2.5}{$\bullet$}} at 20 20   /

\setcoordinatesystem units <1.67pt,1.25pt> point at -175 0
\setplotsymbol ({\scalebox{0.50}{$\bullet$}})
\color{black}
\plot 5 0 20 20 5 40  /
\plot 20 20 50 20 70 20 /
\plot 50 0 50 40  /
\put {$z$} at 35 14   \put {$x$} at 15 5    \put {$y$} at 15 35    
\put {$a$} at 45 10  \put {$b$} at 45 30  \put {$c$} at 60 14  
\color{Sepia}
\multiput {\scalebox{1.5}{$\bullet$}} at 5 0 5 40 50 0 50 20 50 40 70 20  /
\multiput {\scalebox{2.5}{$\bullet$}} at 20 20   /

\color{black}\normalcolor
\endpicture

\caption{A self-avoiding walk, a $3$-star and two double stars.}
\label{FF2}
\end{figure}

In figure \ref{FF2} we show the connectivity of some clusters in the
lattice, from the left, a self-avoiding walk, a 3-star, and then two 
branched structures or trees which may be created by concatenating
two stars using the endpoints of arms.  We shall call these 
\textit{double stars}.  Note that the clusters are rooted at the origin
in a particular node; the self-avoiding walk has first vertex at the
origin, while the star has its central node fixed at the origin.
The double stars are rooted in the central node of one particular
component star.  Noting the location of the root node is important
when counting these, or when joining them, when one has to consider
symmetries in order to exhaust all the possible ways of putting two
or more together.

We proceed by introducing notation to describe the clusters in figure \ref{FF2}.
A self-avoiding walk of length $n$ will be denoted by using square brackets,
so that the walk on the left in figure \ref{FF2} is denoted by $[n]$.  The number 
of such walks is $c_n$.  For example, in the square lattice $c_1=4$, 
$c_2=12$, $c_3=36$ and $c_4=100$. 

The 3-star in figure \ref{FF2} with arms of lengths $x$, $y$ and $z$ will 
be denoted by $[x,y,z]$, and the number of such stars, rooted in the central node, 
will be denoted by $s([x,y,z])$.  For example, one can check that, in the 
square lattice, $s([1,1,1])=4$, $s([2,1,1])=36$ and $s([3,2,1])=520$.  

There are symmetry factors associated with lattice stars.  If a star has
two arms of the same length, then there are two ways of labelling
those arms.  We denote this by $\sigma([x,x,y])=2$
if $x\not=y$. Similarly, $\sigma([x,x,x])=6$ if all three arms have the same length,
but $\sigma([x,y,z])=1$ if $x\not=y\not=z\not=x$.

The double stars in figure \ref{FF2} have clearly defined ``middle'' branches
joining the nodes, and arms attached at the two nodes.  We denote the
middle branch by an extra set of square brackets, so that $[x,y,[z],a,b]$
denotes the double star second from the right in figure \ref{FF2}, 
and $[x,y,[z],a,b,c]$ is the right-most double star.  Notice that $[x,y,0,[z],a,b,0]
= [x,y,[z],a,b]$, etc.

The numbers of branch structures and their symmetry factors are denoted 
similarly to stars.  For example, the number of double stars $[x,y,[z],a,b,c]$
is denoted by $s([x,y,[z],a,b,c])$ and its symmetry factor is
denoted $\sigma([x,y,[z],a,b,c])$.  In all these examples the cluster is rooted
at the fixed large node as shown in figure \ref{FF2} and arms of length zero 
are omitted.  For example, one may check that $s([1,1,[1],1,1])=36$ and 
$\sigma([1,1,[1],1,1])=4$ in the square lattice.  In the cubic lattice,
$s([1,1,[1],1,1])=600$ and $\sigma([1,1,[1],1,1])=4$.

This notation generalizes to combs in a natural way.  First, the backbone
of the comb is composed of segments of (say) lengths $b_1,b_2,\ldots,b_{t+1}$
with $t$ teeth of lengths $u_1,u_2,\ldots,u_t$ substituted along the backbone.
This is denoted by $[[b_1],u_1,[b_2],u_2,\ldots,[b_t],u_t,[b_{t+1}]]$, and notice
that the progression $\ldots,[b_j],u_j,[b_{j+1}],\ldots$ means that a tooth of
length $u_j$ is joined to the node joining backbone segments $j$ and $j{+}1$
of lengths $b_j$ and $b_{j+1}$.  For example, a comb with $3$ teeth of length 
$1$ each, along a backbone with segments all of length $2$, is denoted by 
$[[2],1,[2],1,[2],1,[2]]$, and the $3$-star in figure \ref{FF2} would be $[x,z,y]$,
since it has no backbone. Depending on the choice of a backbone, the third 
and last cases in figure \ref{FF2} are $[x,y,[z],a,b]$ or $[[x],y,[z],a,[b]]$,
and $[x,y,[z],a,b,c]$ or $[x,y,[z],a,b,[c]]$, amongst other possibilities.

This can be generalized to other cases.

\section{A pattern theorem argument}
\label{sec:long}

In this section we show that $\zeta(m_a,m_b) < \kappa$ if  
$m_b$  large but finite and $m_a$ is sufficiently small.

Consider the square lattice and write $U, D, L$ and $R$ for up, down, left and
right steps in the four lattice directions.  Fix a (large) natural number $N=3m{+}5$ 
and consider the pattern $P_N$ given by $R^{m+1}DL^{m+1}DR^{m+1}$.  

Suppose that $m_b<m$.  Since teeth occur on a comb every $m_b$ 
steps of the backbone, the occurrence of $P_N$ in the backbone prevents 
the existence of teeth in a section of the backbone of length $N{+}1$.  
Thus, $P_N$ cannot occur in the backbone of a comb if $m_b<m$.

Denote the number of conformations of a self-avoiding walk not containing 
the pattern $P_N$ by $c_n(\overline{P_N})$ (see equation \BRef{eqn:pattern}).
Then the number of conformations of the backbone is less than or equal to
$c_n(\overline{P_N})$.  Along the backbone each tooth can be appended 
at a node on the backbone with first step in two out of four possible directions.  
This show that, for any fixed $m_b<m$ and $m_a>0$,
\begin{equation}
g(m_a,m_b,t) \le c_{(t+1)m_b}(\overline{P_N})  \left( 2\,c_{m_a}/ 4 \right) ^t,
\label{eqn:patterninequality}
\end{equation}
since there are at most $2\,c_{m_a}/4$ conformations for each tooth. 
Taking logarithms, dividing by $m_b(t+1) + m_at$ and letting $t \to \infty$ gives
\begin{equation}
\zeta(m_a,m_b) 
\le \frac{m_b \kappa_2(\overline{P_N})}{m_a+m_b} 
 + \frac{ \log c_{m_a} -  \log 2}{m_a+m_b},
 \quad\hbox{for any fixed $m_b<m$}.
\label{PLI}
\end{equation}
Since $P_N$ is a Kesten pattern we know that $\kappa_2(\overline{P_N}) < \kappa_2$
\cite{Kesten}.  Putting $m_a=1$, noting that $c_1=4$ and recalling that 
$m_b<m$ for any finite fixed $N$, and that $\kappa_2 > \log 2$, shows that
for any finite $m_b$, $\zeta(1,m_b) < \kappa_2$.

Generalizations of the pattern $P$ to walks in the cubic lattice can be
made, and here one similarly has $\kappa_3(\overline{P_N}) < \kappa_3$.
In this case equation \BRef{PLI} becomes
\begin{equation}
\zeta(m_a,m_b) 
\le \frac{m_b \kappa_2(\overline{P_N})}{m_a+m_b} 
 + \frac{ \log c_{m_a} -  \log (3/2)}{m_a+m_b},
 \quad\hbox{for any fixed $m_b<m$}.
\label{PLI3d}
\end{equation}
since the number of ways a tooth can be attached at a node to the comb is
less than or equal to $(4/6)\,c_{m_a}$.  Putting $m_a=1$ and noting that
$c_1=6$ so that $\log c_1 - \log (3/2) = \log 4 < \log \mu_3$ (see equation
\BRef{lower:mu}), one obtains that, in the cubic lattice, $\zeta(1,m_b) < \kappa_3$.

These results give the following theorem.

\begin{theo}
In the square or cubic lattices, for any finite $m_b>0$, 
$\zeta(1,m_b) < \kappa$. \qed
\label{thm1}
\end{theo}
Since $c_2=12$ and $\log c_2 - \log 2 = \log 6 > \kappa_2$ the above does not
give a similar result for $m_a=2$ in the square lattice.  More generally, 
since $c_n$ is submultiplicative, $\log c_n > n \kappa_2$ in general, so one 
cannot prove, using the above bounds, that $\zeta(m_a,m_b) < \kappa_2$ 
generally, for finite values of $m_a$ and $m_b$ in the square lattice.
Similar observations apply in the hypercubic lattices.  From these considerations
there may be values of $(m_a,m_b)$ where $\zeta(m_a,m_b)$ exceeds $\kappa$.

\begin{table}[t!]
\centering
\caption{Bounds on the comb growth constant $\exp{\zeta(m_a,m_b)}$ 
by equations \BRef{PLI} and \BRef{PLI3d}}
\lineup
\begin{tabular}{lccccccc}
\hline          
 & & Lattice & $m_a=1$ & $m_a=2$ & $m_a=3$ & $m_a=4$ \cr
\hline
\vspace{-2mm}
& & & & & \cr
\multirow{2}{*}{$m_b=1$} 
& & Square & $2.3149^*$ & $2.5239^*$ & $2.6353^\dagger$ & $2.6632$ \cr
& & Cubic   & $4.3740^\dagger$ & $4.5734^\dagger$ & $4.6765^\dagger$ & $4.7088$  \cr \cr
\multirow{2}{*}{$m_b=2$} 
& & Square & $2.4305^*$ & $2.5618^*$ & $2.6440$ & $2.6659$ \cr
& & Cubic   & $4.5062^\dagger$ & $4.6249^\dagger$ & $4.6976$ & $4.7211$   \cr \cr
\multirow{2}{*}{$m_b=3$} 
& & Square & $2.4904^*$ & $2.5849^*$ & $2.6499$ & $2.6678$ \cr
& & Cubic   & $4.5738^\dagger$ & $4.6561^\dagger$ & $4.7117$ & $4.7298$   \cr \cr
\multirow{2}{*}{$m_b=4$} 
& & Square & $2.5271^*$ & $2.6004^*$ & $2.6540$ & $2.6692$ \cr
& & Cubic   & $4.6149^\dagger$ & $4.6770^\dagger$ & $4.7218$ & $4.7364$   \cr \cr
\multirow{2}{*}{$m_b=5$} 
& & Square & $2.5518^*$ & $2.6115^*$ & $2.6572$ & $2.6703$ \cr
& & Cubic   & $4.6425^\dagger$ & $4.6919$ & $4.7294$ & $4.7416$   \cr \cr
\multirow{2}{*}{$m_b=6$} 
& & Square & $2.5696^*$ & $2.6199^*$ & $2.6596$ & $2.6712$ \cr
& & Cubic   & $4.6623^\dagger$ & $4.7032$ & $4.7353$ & $4.7457$   \cr \cr
\multirow{2}{*}{$m_b=7$} 
& & Square & $2.5831^*$ & $2.6264^\dagger$ & $2.6616$ & $2.6719$ \cr
& & Cubic   & $4.6772^\dagger$ & $4.7120$ & $4.7400$ & $4.7490$   \cr 
\hline
\multicolumn{7}{l}{$*$ -- less than lower bound \BRef{lower:mu},
and $\dagger$ -- less than best estimate \BRef{exact:mu}.} \\
\end{tabular}
\label{T1}
\end{table}

In the square lattice more progress can be made using the (exact and rigorous) 
bounds $\log 2.625622 < \kappa_2 < \log 2.679193$ \cite{Jensen2004,PT00}.
Bounding $\kappa_2(\overline{P}) < \log 2.679193$ and 
$\kappa_2 > \log 2.625622$ it follows from equation \BRef{PLI}
that $\zeta(m_a,m_b) < \kappa_2$ when
\begin{equation}
m_b < \frac{m_a \log 2.625622  - \log (c_{m_a}/2)}{
 \log 2.679193 -  \log 2.625622}. 
\label{eqn14}
\end{equation}
Putting $m_a=1$ and $c_{m_a}=4$ then shows that $\zeta(1,m_b) < \kappa_2$
when $m_b\leq 13$.  If $m_a=2$ then $c_{m_a}=12$ and using the above 
shows that $\zeta(2,m_b)<\kappa_2$ when $m_b\leq 6$.  For $m_a\geq 3$
these approximations fail to give useful bounds.  Thus, the result is the following
theorem.

\begin{theo}
In the square lattice, for $1\leq m_b\leq 6$, $\zeta(2,m_b) < \kappa_2$.\hfill \qed
\label{thm2}
\end{theo}

Explicit upper bounds can be obtained using equation 
\BRef{PLI} and the rigorous numerical bounds 
and exact counts for $c_{m_a}$ for small values of $m_a$ \cite{MadrasSlade}.  These are 
shown in table \ref{T1}.  Square lattice upper bounds marked by $*$ are 
below the lower bound $\kappa_2> \log 2.625622$, and so are proofs that 
$\zeta(m_a,m_b) < \kappa_2$.  The bounds for $m_b=1$ and $m_b=2$
are consistent with the statements of theorems \ref{thm1} and \ref{thm2}.

In the cubic lattice an upper bound on $\mu_3$ can be obtained using 
exact enumeration data.  For $n=36$ \cite{SBB11} this gives the 
rigorous bound $\mu_3 \leq 4.7827969$, rounded up to $7$ decimal places. 
Modifying equation \BRef{eqn:patterninequality} to the cubic lattice then gives
\begin{equation}
\hspace{-2cm}
\zeta(m_a,m_b) 
\le \frac{m_b \kappa_3(\overline{P_N})}{m_a+m_b} 
 + \frac{ \log c_{m_a} -  \log (3/2)}{m_a+m_b}
< \frac{m_b \log 4.7827969 + \log c_{m_a} - \log (3/2) }{m_a+m_b} ,
\label{PLI3d}
\end{equation}
using $\kappa_3(\overline{P}) \leq \kappa_3 \leq \log 4.7827969$.  This gives, 
for example, $\zeta(1,1) \leq \log 4.3740$ which is well below the best numerical 
estimate for $\log \mu_3$ in equation \BRef{exact:mu}, but still above the 
rigorous lower bound on $\log \mu_3$ in equation \BRef{lower:mu}.  More 
cubic lattice bounds are shown in table \ref{T1}.

\section{Upper bounds on the growth rate from exact counts of self-avoiding walks}
\label{sec:growthrate}

In this section we investigate the situation when the teeth 
are short, and show that there are cases where the growth constant 
is strictly less than that of self-avoiding walks.

\subsection{Short teeth}
\label{sec:short}
The next approach that we develop to investigate the case of short teeth 
makes use of an idea due to Lipson \cite{Lipson1993}.  The essential feature
is the existence of vertices of degree 3 that limit the number of ways of 
embedding the short teeth.  To obtain an upper bound we regard the teeth and the 
backbone as independent, and the teeth as independent of one another.  This gives 
the inequality
\begin{equation}
g(m_a,m_b,t) \le c_{(t+1)m_b} \left( (\gamma{-}2)c_{m_a}/ \gamma\right) ^t ,
\label{eqn:maininequality}
\end{equation}
where $\gamma$ is the coordination number of the lattice.  The first factor reflects the 
fact that the backbone of the comb consists of $t{+}1$ segments, each of length 
$m_b$.  The factor of $(\gamma{-}2)/\gamma$ in brackets is due to the fact that each 
tooth can start in $\gamma{-}2$ ways so the number of walks making up the tooth is 
reduced by the factor $(\gamma{-}2)/\gamma$. If we take logarithms, divide by the total number 
of edges, $tm_a + (t{+}1)m_b$, and let $t \to \infty$ we have the inequality
\begin{equation}
\zeta(m_a,m_b) \le \frac{m_b \log \mu}{m_a+m_b} 
  + \frac{\log ((\gamma{-}2)/\gamma) + \log c_{m_a}}{m_a+m_b}.
\label{eqn:mainlimitinginequality}
\end{equation}

\subsubsection{The square lattice}
\label{sec:square}

For the square lattice (\ref{eqn:mainlimitinginequality}) gives 
\begin{equation}
\zeta(m_a,m_b) \le \frac{ m_b \log\mu_2 }{m_a+m_b} + \frac{\log(c_{m_a}/2)}{m_a+m_b},
\end{equation}
and $\zeta(m_a,m_b) < \log \mu_2$ if $\mu_2^{m_a} > c_{m_a}/2$.  
From \cite{Jensen2004} we know that $\mu_2 \ge 2.625622$
and for $m_a=1$, $c_1/2=2$.  Thus $\zeta(1,m_b) < \kappa_2$ for all
$m_b\geq 1$ as seen in theorem \ref{thm1}.  If $m_a=2$, then 
$\mu_2^2 \ge 6.89 > c_2/2=6$ and $\zeta(2,m_b)<\kappa_2$ for all 
$m_b\geq 1$, improving theorem \ref{thm2}. Continuing to 
$m_a=3$ gives $\mu_2^3 \ge 18.1 > c_3/2=18$ and thus
$\zeta(3,m_b)<\kappa_2$ for all $m_b\geq 1$.  However, if
$m_a=4$ then $\mu_2^4>47.53$ while $c_4/2=50$, and so this 
breaks down.  This gives the following theorem. 

\begin{theo}
In the square lattice, if $1\leq m_a\leq 3$, then $\zeta(m_a,m_b) < \kappa_2$ for all finite $m_b\geq 1$. \qed
\label{thm3}
\end{theo}

To obtain rigorous numerical bounds this again reduces
to the results in table \ref{T1}, in particular since the best numerical
bound is $\kappa_2 <\log 2.679193$ \cite{PT00} in theorem \ref{thm3}.

In the cubic lattice, using the lower bounds in equation \BRef{lower:mu}
\cite{FisherSykes} in a similar way, we see that $\zeta(m_a,m_b) < \log \mu_3$
if $\mu_3^{m_a} > 2\,c_{m_a}/3$, which is true if $4.225^{m_a}> 2\,c_{m_a}/3$.
This is the case for $m_a=1$ and any finite $m_b\geq 1$.  It follows
that $\zeta(1,m_b) < \log \mu_3$ for all finite $m_b\geq 1$ (as already seen
in theorem \ref{thm1}).

\subsubsection{The hexagonal lattice}
\label{sec:hexagonal}

For the hexagonal lattice the coordination number is 3 and we know \cite{Smirnov} 
the exact value of $\mu = \mu_H = \sqrt{2+\sqrt{2}}$ rigorously, and we 
define $\kappa_H = 
\log \mu_H$.  Equation (\ref{eqn:mainlimitinginequality}) then gives
\begin{equation}
\zeta(m_a,m_b) \le \frac{ m_b \kappa_H }{m_a+m_b} + \frac{\log(c_{m_a}/3)}{m_a+m_b}.
\end{equation}
Rearranging gives
\begin{equation}
\zeta(m_a,m_b) \le   \kappa_H - \left( \frac{m_a \kappa_H - \log (c_{m_a}/3)}{m_a+m_b}    \right).
\label{eqn:hexagonalinequality}
\end{equation}
$\zeta(m_a,m_b) < \kappa_H$ if $\mu_H^{m_a} = (\sqrt{2+\sqrt{2}})^{m_a} 
> c_{m_a}/3$.   Exact counts of $c_n$ for the hexagonal lattice are available to
$n=105$\footnote{Jensen: https://oeis.org/A001668} \cite{GuttmannSykes,Jensen2004}.
Examining these shows that $\zeta(m_a,m_b) < \kappa_H$ for $m_a \leq 15$
(where $c_n=29892$). This gives the following theorem.

\begin{theo}
In the hexagonal lattice, $\zeta(m_a,m_b) < \kappa_H$ for all $1\leq m_a \leq 15$
and for all finite $m_b\geq 1$. \qed
\end{theo}


\subsubsection{Some other lattices}
\label{sec:otherlattices}

In the triangular lattice the growth constant has bounds
$4.118935 \leq \mu_T \leq 4.25152$ \cite{Jensen2004,Alm2005}.  Defining
$\kappa_T = \log \mu_T$, equation (\ref{eqn:mainlimitinginequality}) gives 
\begin{equation}
\zeta(m_a,m_b) \le \frac{ m_b \kappa_T }{m_a+m_b} 
 + \frac{\log(2\,c_{m_a}/3)}{m_a+m_b},
\label{eqn22}
\end{equation}
and $\zeta(m_a,m_b) < \kappa_T$ if $m_a \kappa_T > \log( 2\, c_{m_a}/3)$.
This shows that for $m_a=1$, $\zeta(1,m_b)<\kappa_T$.

\begin{theo}
In the triangular lattice, $\zeta(1,m_b) < \kappa_T$ for all finite $m_b\geq 1$. \qed
\label{tri}
\end{theo}

Notice that the numerical bounds above can be used with equation \BRef{eqn22}
to find explicit bounds, but that none of these are better than the result in
theorem \ref{tri}.

Bounds on the growth constant $\mu_K$ in the Kagome lattice are 
$2.548497 < \mu_K < 2.590301$ \cite{Jensen2004,Guttmann2004}.  The 
coordination number is $\gamma=4$.  Defining $\kappa_K = \log \mu_K$ and 
using exact counts \cite{Alm2005} show that if $m_a \kappa_K > \log c_{m_a}/2$ 
then $\zeta(m_a,m_b) < \kappa_K$ if $m_a\leq 3$ and for all finite $m_b\geq 1$.

\begin{theo}
In the Kagome lattice, $\zeta(m_a,m_b) < \kappa_K$ if $1\leq m_a\leq 3$ 
and for all finite $m_b\geq 1$. \qed
\end{theo}

\section{Upper bounds derived using lattice stars and other lattice clusters}
\label{sec:stars}

In the last section combs were constructed from several self-avoiding walks.  
Self-avoidance was relaxed between the backbone and teeth, as well as between
teeth, to obtain upper bounds on $\zeta(m_a,m_b)$.  An alternative approach
would be to decompose the comb into other sub-units (that are self-avoiding 
clusters in the lattice). While these are not mutually avoiding, they do 
give useful bounds in many cases. Examples of such clusters are shown 
in figure \ref{FF2}.  In order to use these clusters to determine bounds, 
exact counts will be needed.  For this we implemented a back-tracking 
algorithm coded to enumerate clusters of various fixed connectivities and 
branches of given sizes.  In tables \ref{E1} and \ref{E2} some counts 
for lattice $3$-stars in the square and cubic lattices are shown.

\subsection{Creating combs by concatenating walks and lattice $3$-stars}

In the case of combs there are natural decompositions into sub-units which
are combinations of lattice stars and walks, or even larger sub-clusters.
By counting the number of conformations of the clusters, upper bounds on the
number of conformations of the combs are obtained (since the sub-units, while
self-avoiding, are not mutually avoiding).

In what follows we limit the discussion to the hypercubic lattice.  Similar
results can be obtained in other lattices.

\renewcommand{\arraystretch}{1.33}
\begin{table}[h!]
\centering
\caption{Square lattice self-avoiding walk and $3$-star counts }
\lineup
\begin{tabular}{l|r@{}*{9}{r}}
\hline          
 \scalebox{0.9}{$n$}  &
 \scalebox{0.9}{$c_n$} & &
 \scalebox{0.9}{$s([n,1,1])$} & 
 \scalebox{0.9}{$s([n,2,2])$} & 
 \scalebox{0.9}{$s([n,3,3])$} & 
 \scalebox{0.9}{$s([n,4,4])$} & 
 \scalebox{0.9}{$s([n,5,5])$} & 
 \scalebox{0.9}{$s([n,6,6])$} & 
 \scalebox{0.9}{$s([n,n,n])$}  \cr
\hline
1  & 4 & & 4 & 100 & 652 & 5252 & 36836 & 281132 &  4  \cr
2  & 12 & & 36 & 84 & 1700 & 12676 & 91316 & 664324 &  84 \cr
3  & 36 & & 92 & 668 & 1380 & 31692 & 216452 & 1602428 &  1380 \cr
4  & 100 & & 268 & 1820 & 11588 & 28164 & 587796 & 4211372 &  28164 \cr
5  & 284 & & 716 & 4988 & 30244 & 224620 & 504084 & 10961516 &  504084 \cr
6  & 780 & & 2020 & 13516 & 83404 & 600892 & 4093036 & 9675548 &  9675548 \cr
\hline
\end{tabular}
\label{E1}
\end{table}

\renewcommand{\arraystretch}{1.33}
\begin{table}[h!]
\centering
\caption{Cubic lattice self-avoiding walk and $3$-star counts }
\lineup
\resizebox{\columnwidth}{!}{%
\begin{tabular}{l|r@{}*{9}{r}}
\hline          
 \scalebox{0.9}{$n$}  &
 \scalebox{0.9}{$c_n$} & &
 \scalebox{0.9}{$s([n,1,1])$} & 
 \scalebox{0.9}{$s([n,2,2])$} & 
 \scalebox{0.9}{$s([n,3,3])$} & 
 \scalebox{0.9}{$s([n,4,4])$} & 
 \scalebox{0.9}{$s([n,5,5])$} & 
 \scalebox{0.9}{$s([n,6,6])$} & 
 \scalebox{0.9}{$s([n,n,n])$}  \cr
\hline
1  & 6 & & 20 & 1452 & 31596 & 734196 & 16174908 & 368476164 & 20  \cr
2  & 30 & & 300 & 2260 &149052 & 3350604 & 74400420 & 1665999876 & 2260 \cr
3  & 150 & & 1404 & 32052 & 227172 & 15434316 & 336611868 & 7572405156 & 227172 \cr
4  & 726 & & 6876 & 152052 & 3257412 & 24159620 & 1588051596 & 35328582708 & 24159620 \cr
5  & 3534 & & 32436 & 722988 & 15219372 & 340216500 & 2456804660 & 164468765436 & 2456804660 \cr
6  & 16926 & & 156588 & 3432684 & 72597636 & 1605227796 & 34882592748 & 257495456028 & 257495456028 \cr
\hline
\end{tabular}
}  
\label{E2}
\end{table}

\begin{figure}[h!]
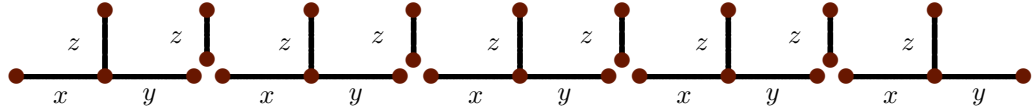

    \beginpicture
\setcoordinatesystem units <1.67pt,1.25pt>
\setplotarea x from 0 to 270, y from -10 to 30

\setplotsymbol ({\scalebox{0.50}{$\bullet$}})
\color{black}
\plot 20 0 60 0 /
\plot 40 0 40 20 /
\plot 63 5 63 20 /  \put {$z$} at 56 12 
\put {$z$} at 33 10   \put {$x$} at 30 -6    \put {$y$} at 50 -6    
\color{Sepia}
\multiput {\scalebox{1.5}{$\bullet$}} at 20 0 40 0 40 20 60 0 63 5 63 20  /

\setcoordinatesystem units <1.67pt,1.25pt> point at -2 0
\multiput {
\beginpicture
\setplotsymbol ({\scalebox{0.50}{$\bullet$}})
\color{black}
\plot  0 0 40 0 /
\plot  20 0 20 20 /
\plot  43 5 43 20 / \put {$z$} at 35 12
\put {$x$} at 10 -6   \put {$y$} at 30 -6    \put {$z$} at 14 10    
\color{Sepia}
\multiput {\scalebox{1.5}{$\bullet$}} at 0 0 20 0 20 20 40 0 43 5 43 20  / 
\endpicture
} at 85 7 132 7 179 7 /

\setplotsymbol ({\scalebox{0.50}{$\bullet$}})
\color{black}
\plot 205 0 245 0 /
\plot 225 0 225 20 /
\put {$z$} at 219 10   \put {$x$} at 215 -6    \put {$y$} at 235 -6    
\color{Sepia}
\multiput {\scalebox{1.5}{$\bullet$}} at 205 0 245 0 225 0 225 20  /

\color{black}\normalcolor
\endpicture

\caption{The decomposition of a comb into a series of $3$-stars and walks.}
\label{F2}
\end{figure}

As an example, in figure \ref{F2} a comb is decomposed into a sequence of
$3$-stars with arms of lengths $(x,y,z)$ (these are denoted by $[x,z,y]$) 
alternating with walks of length $z$.  Putting $z=m_a$ and $x=y=m_b$ shows 
that this is a decomposition of a comb into $3$-stars with arms of lengths 
$(m_a,m_b,m_b)$ and walks of length $m_a$. The number of such stars rooted 
at the origin in the central node, is denoted $s([m_b,m_a,m_b])$.  

In order to determine an upper bound using the decomposition in figure
\ref{F2}, careful accounting of symmetry factors is needed. 

Concatenating lattice stars at the endpoints of arms as shown in figure \ref{F2} 
in effect moves the root from the central node to the endpoints of arms.  If there 
are several arms which have the same length, then this introduces a symmetry factor 
in the concatenation.  In the particular case that $3$-stars are concatenated as in
figure \ref{F2}, proceed as follows.  If $x\not=y\not=z\not=x$ then
the number of $3$-stars rooted at the endpoint of an arm of length $x$ is
$s([x,y,z])$.  If $x=y$, or $x=z$, but $y\not=z$, then there are two 
choices for the root at the endpoint of an arm of length $x$, so the number
is then $2\,s([x,x,z])$.  If $x=y=z$, then the number is $3\,s([x,x,x])$, by a 
similar argument. 

With the above in mind, put $x=y=m_b$ and $z=m_a$ in figure \ref{F2}.
There are $2\,s([m_a,m_b,m_b])$ choices for the left-most star with the endpoint
of an arm of length $m_b$ rooted at the origin.  There are, similarly,
$2\,s([m_a,m_b,m_b])$ choices for the second star.  In addition, concatenating
the stars at the endpoint of arms of length $m_b$ requires that the last steps
of these arms have different orientations.  This reduces the number of
choices for the second star by a factor $(2d{-}1)/(2d))$ to 
$((2d{-}1)/(2d))\times 2\,s(m_a,m_b,m_b)$.  

To attach a tooth, we note that there are $((2d{-}2)/(2d))\times c_{m_a}$ 
ways of selecting the tooth, since two directions are already taken by
the stars where they are joined together.  This remains true when concatenating 
the subsequent stars.  Thus, if the comb has $t$ teeth \emph{and $m_a\not=m_b$}, 
then an upper bound on the number of combs is obtained (since the component 
walks and stars are not mutually avoiding):
\begin{eqnarray*}
g(m_a,m_b,t) &\leq& (2\,s([m_a,m_b,m_b]))\times
\left( (2d{-}1)/(2d) \times 2\,s([m_a,m_b,m_b])\right)^{\lceil t/2\rceil-1} \\
& &\quad \times
\left( (2d{-}2)/(2d) \times\,c_{m_a} \right)^{\lfloor t/2 \rfloor} .
\end{eqnarray*}
Taking logarithms, dividing by $tm_a+(t{+}1)m_b$, and taking $t\to\infty$ gives
\begin{equation}
\hspace{-2.0cm}
\zeta(m_a,m_b) \leq
\left( \log((2d{-}1)(d{-}1)/d^2) + \log s([m_a,m_b,m_b]) 
        + \log c_{m_a} \right)/(2(m_a{+}m_b)) .
\end{equation}
In the square lattice, this gives
\begin{equation}
\zeta(m_a,m_b) \leq ( \log(3/4) + \log s([m_a,m_b,m_b]) + \log c_{m_a} )/(2(m_a{+}m_b)) .
\label{eqn26}
\end{equation}

In the case that $m_a=m_b$ a more careful accounting of joining a star to the previous
star gives the inequality
\begin{equation}
\hspace{-2.0cm}
\zeta(m_b,m_b) \leq
\left( \log(3(2d{-}1)(d{-}1)/(d^2)) + \log s([m_b,m_b,m_b]) 
        + \log c_{m_b} \right)/(2(m_b{+}m_b)) .
\end{equation}
In the square lattice this gives
\begin{equation}
\zeta(m_b,m_b) \leq (\log(9/4) + \log s([m_b,m_b,m_b]) + \log c_{m_b} )/(4m_b) .
\label{eqn27}
\end{equation}
We state these results as theorems.

\begin{theo}
If $m_a\not=m_b$ then the following bounds hold for $\zeta(m_a,m_b)$ and $\zeta(m_b,m_b)$:
\begin{eqnarray*}
& \hspace{-1cm}
\zeta(m_a,m_b) \leq ( \log((2d{-}1)(d{-}1)/d^2) + \log s([m_a,m_b,m_b]) + \log c_{m_a} )/(2(m_a{+}m_b)) ,\cr
&  \hspace{-1cm}
\zeta(m_b,m_b) \leq (\log(3(2d{-}1)(d{-}1)/d^2) + \log s([m_b,m_b,m_b]) + \log c_{m_b} )/(4m_b) . 
\end{eqnarray*}
These inequalities are true in the hypercubic lattice.  \hfill \qed
\label{thm7}
\end{theo}

\begin{table}[h!]
\centering
\caption{Bounds on the comb growth constant $\exp{\zeta(m_a,m_b)}$ by theorem \ref{thm7}}
\lineup
\begin{tabular}{lcccccccc}
\hline          
 & & Lattice & $m_a=1$ & $m_a=2$ & $m_a=3$ & $m_a=4$ & $m_a=5$ & $m_a=6$ \cr
\hline
\vspace{-2mm}
& & & & & \cr
\multirow{2}{*}{$m_b=1$} 
& & Square & $ 2.4495^*$ & $ 2.6208^*$ & $ 2.6571$ & $ 2.6936$ & $ 2.7036$ & $ 2.7149$ \cr         
& & Cubic & $ 4.4722^\dagger$ & $ 4.6416^\dagger$ & $ 4.6898$ & $ 4.7251$ & $ 4.7351$ & $ 4.7464$ \cr \cr
\multirow{2}{*}{$m_b=2$} 
& & Square & $ 2.5874^*$ & $ 2.6270^\dagger$ & $ 2.6645$ & $ 2.6788$ & $ 2.6944$ & $ 2.6986$ \cr
& & Cubic & $ 4.6165^\dagger$ & $ 4.6695^\dagger$ & $ 4.7073$ & $ 4.7213$ & $ 4.7340$ & $ 4.7383$ \cr \cr
\multirow{2}{*}{$m_b=3$} 
& & Square & $ 2.5789^*$ & $ 2.6211^*$ & $ 2.6345^\dagger$ & $ 2.6560$ & $ 2.6642$ & $ 2.6739$ \cr
& & Cubic & $ 4.6286^\dagger$ & $ 4.6733^\dagger$ & $ 4.6912$ & $ 4.7079$ & $ 4.7155$ & $ 4.7226$ \cr \cr
\multirow{2}{*}{$m_b=4$} 
& & Square & $ 2.6288^\dagger$ & $ 2.6390$ & $ 2.6530$ & $ 2.6615$ & $ 2.6709$ & $ 2.6749$ \cr
& & Cubic & $ 4.6663^\dagger$ & $ 4.6846^\dagger$ & $ 4.7008$ & $ 4.7091$ & $ 4.7169$ & $ 4.7206$ \cr \cr
\multirow{2}{*}{$m_b=5$} 
& & Square & $ 2.6320^\dagger$ & $ 2.6454$ & $ 2.6481$ & $ 2.6588$ & $ 2.6632$ & $ 2.6689$ \cr
& & Cubic & $ 4.6709^\dagger$ & $ 4.6885$ & $ 4.6969$ & $ 4.7059$ & $ 4.7106$ & $ 4.7172$ \cr \cr
\multirow{2}{*}{$m_b=6$} 
& & Square & $ 2.6503$ & $ 2.6518$ & $ 2.6561$ & $ 2.6606$ & $ 2.6659$ & $ 2.6684$ \cr
& & Cubic & $ 4.6853$ & $ 4.6940$ & $ 4.7020$ & $ 4.7070$ & $ 4.7118$ & $ 4.7144$ \cr
\hline
\multicolumn{8}{l}{$*$ -- less than lower bound \BRef{lower:mu},
and $\dagger$ -- less than best estimate \BRef{exact:mu}.} \\
\end{tabular}
\label{T3}
\end{table}

Using theorem \ref{thm7} and noting that $s([1,1,1])=c_1=4$ in the square lattice, 
$\zeta(1,1) \leq \log \sqrt{6} < \log 2.4495$ (where the bound is rounded up in the 
$4$-th decimal place).  This improves the bound in table \ref{T1}.  In table \ref{T3} 
additional bounds are shown on $\zeta(m_a,m_b)$ in the square and cubic lattices,
for $1\leq m_a,m_b \leq 6$, by using the counts in tables \ref{E1} and \ref{E2}
in theorem \ref{thm7}.

We note that the bounds in table \ref{T1} are due to self-avoiding walk bounds
on both the backbone and teeth of the comb.  In contrast the construction in
figure \ref{F2} takes into account the local reduction in conformations when
the stars and walks are concatenated, reducing the conformational degrees
of freedom of the backbone.  However, intersections between the components may
occur, so that the backbone is not a self-avoiding walk.  Generally, the quality
of the bounds in theorem \ref{thm7} should improve with $m_b$, since larger stars
will result in longer segments along the backbone being self-avoiding.
Indeed,  equation \BRef{PLI}
(using the upper bound $\kappa_2 < \log 2.679193$ \cite{PT00}) gives
the (rigorous) bound $\zeta(10,10) < \log 2.6989$ while
fixing $m_a=m_b=10$ in theorem \ref{thm7} gives the better bound 
$\zeta(10,10) < \log 2.6682$ (solely relying on the exact counts 
$s([10,10,10])=1126242913156$ and $c_{10}=44100$).  This is smaller
than the bound determined on $\zeta(6,6)$ in table \ref{T3}.

\subsection{Concatenating asymmetric lattice stars}

\begin{figure}[h!]
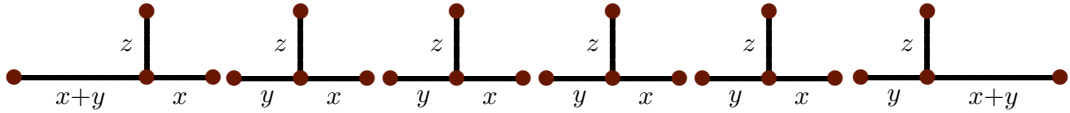

    \beginpicture
\setcoordinatesystem units <1.25pt,1.25pt>
\setplotarea x from -10 to 270, y from -10 to 30

\setplotsymbol ({\scalebox{0.50}{$\bullet$}})
\color{black}
\plot 0 0 60 0 /
\plot 40 0 40 20 /
\put {$z$} at 34 10   \put {$x{+}y$} at 20 -6    \put {$x$} at 50 -6    
\color{Sepia}
\multiput {\scalebox{1.5}{$\bullet$}} at 0 0 40 0 40 20 60 0  /

\multiput {
\beginpicture
\setplotsymbol ({\scalebox{0.50}{$\bullet$}})
\color{black}
\plot  0 0 40 0 /
\plot  20 0 20 20 /
\put {$y$} at 10 -6   \put {$x$} at 30 -6    \put {$z$} at 14 10    
\color{Sepia}
\multiput {\scalebox{1.5}{$\bullet$}} at 0 0 20 0 20 20 40 0  / 
\endpicture
} at 85 7 132 7 179 7 226 7 /

\setplotsymbol ({\scalebox{0.50}{$\bullet$}})
\color{black}
\plot 255 0 315 0 /
\plot 275 0 275 20 /
\put {$z$} at 269 10   \put {$x{+}y$} at 295 -6    \put {$y$} at 265 -6    
\color{Sepia}
\multiput {\scalebox{1.5}{$\bullet$}} at 255 0 275 0 275 20 315 0  /

\color{black}\normalcolor
\endpicture

\caption{Constructing a lattice comb by concatenating a sequence of
lattice $3$-stars.}
\label{F3}
\end{figure}

\begin{figure}[h!]
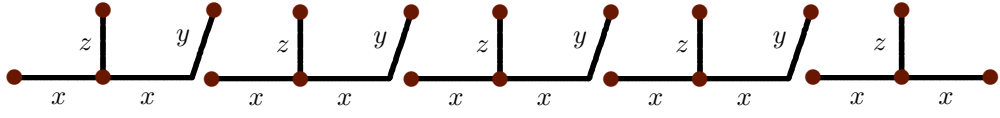

    \beginpicture
\setcoordinatesystem units <1.67pt,1.25pt>
\setplotarea x from 0 to 270, y from -10 to 30

\setplotsymbol ({\scalebox{0.50}{$\bullet$}})
\color{black}
\plot 20 0 60 0 /
\plot 40 0 40 20 /
\plot 60 0 65 20 /  \put {$y$} at 58 12 
\put {$z$} at 36 10   \put {$x$} at 30 -6    \put {$x$} at 50 -6    
\color{Sepia}
\multiput {\scalebox{1.5}{$\bullet$}} at 20 0 40 0 40 20 65 20  /

\multiput {
\beginpicture
\setplotsymbol ({\scalebox{0.50}{$\bullet$}})
\color{black}
\plot  0 0 40 0 /
\plot  20 0 20 20 /
\plot  40 0 45 20 / \put {$y$} at 38 12
\put {$x$} at 10 -6   \put {$x$} at 30 -6    \put {$z$} at 16 10    
\color{Sepia}
\multiput {\scalebox{1.5}{$\bullet$}} at 0 0 20 0 20 20  45 20  / 
\endpicture
} at 86 7 131 7 176 7 /

\setplotsymbol ({\scalebox{0.50}{$\bullet$}})
\color{black}
\plot 200 0 240 0 /
\plot 220 0 220 20 /
\put {$z$} at 215 10   \put {$x$} at 210 -6    \put {$x$} at 230 -6    
\color{Sepia}
\multiput {\scalebox{1.5}{$\bullet$}} at 200 0 240 0 220 0 220 20  /

\color{black}\normalcolor
\endpicture

\caption{Constructing a lattice comb by concatenating $3$-stars.}
\label{F4}
\end{figure}

Similar to the approach above, it is possible to use lattice stars in other ways
to determine upper bounds on  $\zeta(m_a,m_b)$.  In this section we focus 
on the constructions shown in figures \ref{F3} and \ref{F4}.

The general methods here are the same as in the last section, and so we
simply state the upper bound as a theorem.   The factors arising from
symmetry considerations are accounted for by defining a function
$\varsigma([x,y,z])$ on stars $[x,y,z]$ by putting $\varsigma([x,x,x])=6$, 
$\varsigma([x,x,z])=\varsigma([x,z,x])=\varsigma([z,x,x])=2$ if $z\not=x$,
and $\varsigma([x,y,z])=1$ otherwise, for $x,y,z\in \mathbb{N}$.

The construction in figure \ref{F3} gives the following theorem.

\begin{theo}
$ \displaystyle \zeta(z,x{+}y) \leq \left(
\log ((2d{-}1)\varsigma([x,y,z])/2d ) + \log s([x,y,z])
\right) / (x{+}y{+}z).$ \hfill \qed
\label{thm8}
\end{theo}

The result in theorem \ref{thm8} is not very useful in determining bounds 
on $\zeta(m_a,m_b)$.  Instead, a decomposition as shown in figure \ref{F4}
consisting of larger 3-star components may be considered.  Denoting the
growth constant of combs with $t$ teeth of alternating lenghts $z$ and $y$
and backbone segments of length $x$ by $\xi(x,y,z)$, the following
lemma is obtained from figure \ref{F4}.

\begin{lemm}
$ \displaystyle \xi(x,y,z) \leq \left(
\log ((d{-}1)\varsigma([x,x{+}y,z])/d ) + \log s([x,x{+}y,z])
\right) / (2x{+}y{+}z).$ \hfill \qed
\label{lemm1}
\end{lemm}

By lemma \ref{lemm1}:

\begin{theo}  
Since $\zeta(m_a,m_b) = \xi(m_b,m_a,m_a)$, \\
$ \zeta(m_a,m_b)  
  \leq (\log ((d{-}1)\varsigma([m_a,m_b,m_a{+}m_b])/d)
   + \log s([m_a,m_b,m_a{+}m_b])) / 2(m_a{+}m_b)$.\hfill \qed 
\label{thm9}
\end{theo} 

This shows, for the square lattice, that
\begin{equation}
\hspace{-2.5cm}
\zeta(1,1) \leq (\log(\varsigma([1,1,2])/2) + \log s([1,1,2]) )/4 = (\log 1+\log 36)/4 = 
\log \sqrt{6} < \log 2.4495 . 
\end{equation}
This is the same result as obtained in theorem \ref{thm7}.  In table \ref{T5} we 
give square and cubic lattice bounds due to theorem \ref{thm9} on
$\zeta(m_a,m_b)$ for $1\leq m_a,m_b\leq 6$ (using the data in table \ref{A1}).  

\begin{table}[h!]
\centering
\caption{Bounds on the comb growth constant $\exp{\zeta(m_a,m_b)}$ by theorem \ref{thm9}}
\lineup
\begin{tabular}{l@{}*{3}{c}ccccc}
\hline          
 & & Lattice & $m_a=1$ & $m_a=2$ & $m_a=3$ & $m_a=4$ & $m_a=5$ & $m_a=6$ \cr 
\hline
\vspace{-2mm}
& & & & & \cr
\multirow{2}{*}{$m_b=1$} 
& & Square & $ 2.4495^*$ & $ 2.5264^*$ & $ 2.5641^*$ & $ 2.5980^*$ & $ 2.6150^*$ & $ 2.6281^\dagger$ \cr
& & Cubic & $ 4.4722^\dagger$ & $ 4.5696^\dagger$ & $ 4.6099^\dagger$ & $ 4.6382^\dagger$ & $ 4.6535^\dagger$ & $ 4.6655^\dagger$  \cr \cr
\multirow{2}{*}{$m_b=2$} 
& & Square & $ 2.5264^*$ & $ 2.5557^*$ & $ 2.5702^*$ & $ 2.5929^*$ & $ 2.6054^*$ & $ 2.6159^*$ \cr
& & Cubic & $ 4.5696^\dagger$ & $ 4.6065^\dagger$ & $ 4.6233^\dagger$ & $ 4.6413^\dagger$ & $ 4.6515^\dagger$ & $ 4.6607^\dagger$  \cr \cr
\multirow{2}{*}{$m_b=3$} 
& & Square & $ 2.5641^*$ & $ 2.5699^*$ & $ 2.5710^*$ & $ 2.5882^*$ & $ 2.5959^*$ & $ 2.6055^*$ \cr
& & Cubic & $ 4.6099^\dagger$ & $ 4.6233^\dagger$ & $ 4.6291^\dagger$ & $ 4.6416^\dagger$ & $ 4.6481^\dagger$ & $ 4.6558^\dagger$  \cr \cr
\multirow{2}{*}{$m_b=4$} 
& & Square & $ 2.5980^*$ & $ 2.5929^*$ & $ 2.5882^*$ & $ 2.5987^*$ & $ 2.6035^*$ & $ 2.6099^*$ \cr
& & Cubic & $ 4.6382^\dagger$ & $ 4.6413^\dagger$ & $ 4.6416^\dagger$ & $ 4.6493^\dagger$ & $ 4.6534^\dagger$ & $ $  \cr \cr
\multirow{2}{*}{$m_b=5$} 
& & Square & $ 2.6150^*$ & $ 2.6054^*$ & $ 2.5959^*$ & $ 2.6035^*$ & $ 2.6054^*$ & $ 2.6107^*$ \cr
& & Cubic & $ 4.6535^\dagger$ & $ 4.6515^\dagger$ & $ 4.6481^\dagger$ & $ 4.6534^\dagger$ & $ $ & $ $  \cr \cr
\multirow{2}{*}{$m_b=6$} 
& & Square & $ 2.6281^\dagger$ & $ 2.6159^*$ & $ 2.6055^*$ & $ 2.6099^*$ & $ 2.6107^*$ & $ 2.6141^*$ \cr
& & Cubic & $ 4.6655^\dagger$ & $ 4.6607^\dagger$ & $ 4.6558^\dagger$ & $ $ & $ $ & $ $  \cr
\hline
\multicolumn{8}{l}{$*$ -- less than lower bound \BRef{lower:mu},
and $\dagger$ -- less than best estimate \BRef{exact:mu}.} \\
\end{tabular}
\label{T5}
\end{table}

\begin{figure}[h!]
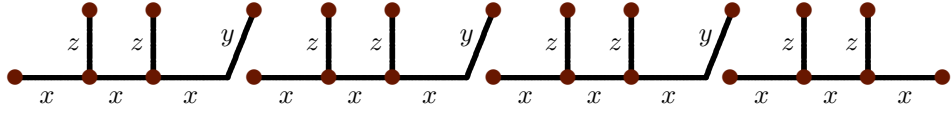

    \beginpicture
\setcoordinatesystem units <2pt,1.25pt>
\setplotarea x from 0 to 270, y from -10 to 30

\put {$ $} at 0 0

\multiput {
\beginpicture
\setplotsymbol ({\scalebox{0.50}{$\bullet$}})
\color{black}
\plot 20 0 60 0 /
\plot 34 0 34 20 /
\plot 46 0 46 20 /
\plot 60 0 65 20 /  \put {$y$} at 60 12 
\put {$z$} at 31 10   \put {$z$} at 43 10   
\put {$x$} at 26 -6    \put {$x$} at 39 -6    \put {$x$} at 53 -6    
\color{Sepia}
\multiput {\scalebox{1.5}{$\bullet$}} at 20 0 34 0 34 20 46 0 46 20 65 20  /
\endpicture}
at 40 10  85 10 130 10 /  

\multiput {
\beginpicture
\setplotsymbol ({\scalebox{0.50}{$\bullet$}})
\color{black}
\plot 20 0 60 0 /
\plot 34 0 34 20 /
\plot 46 0 46 20 /
\put {$z$} at 31 10   \put {$z$} at 43 10   
\put {$x$} at 26 -6    \put {$x$} at 39 -6    \put {$x$} at 53 -6    
\color{Sepia}
\multiput {\scalebox{1.5}{$\bullet$}} at 20 0 34 0 34 20 46 0 46 20 60 0  /
\endpicture}
at 172 10  /  

\color{black}\normalcolor
\endpicture

\caption{Concatening larger clusters to bound the number of combs.}
\label{F5}
\end{figure}

Better bounds can be obtained by concatenating larger lattice clusters 
to create the comb.  If $\chi(x,y,z)$ is the connective constant of the comb constructed
in figure \ref{F5}, then
\begin{equation}
\hspace{-2cm}
\chi(x,y,z) \leq (\log ((d{-}1)\varsigma([x,z,[x],x{+}y,z])/d) +
 \log s([x,z,[x],x{+}y,z]) )/(3x{+}y{+}2z) .
\end{equation}
This gives the following bound.

\begin{theo}
Since $\zeta(m_a,m_b) = \chi(m_b,m_a,m_a)$, \\
$\scalebox{0.9}{$\displaystyle
\zeta(m_a,m_b) \leq
(\log( (d{-}1)\varsigma([m_b,m_a,[m_b],m_b{+}m_a,m_a])/d)
  + \log s([m_b,m_a,[m_b],m_b{+}m_a,m_a]))/3(m_a{+}m_b) $}$. \qed
\label{thm10}
\end{theo}

Using theorem \ref{thm10}, and noting that in the square lattice
$s([1,1,[1],2,1])=184$ and $\varsigma([1,1,[1],2,1])=2$,
$\zeta(1,1) \leq (\log(2/2)+\log 184)/6 < \log 2.3849$, improving on the bound
derived in tables \ref{T3} and \ref{T5}.  In the cubic lattice $s([1,1,[1],2,1])= 5616$ 
and $\varsigma([1,1,[1],2,1])=2$ so that
$\zeta(1,1) \leq (\log(2\times 2/3) + \log 5616)/6 < \log 4.4232$.  This is smaller
than the bound in table \ref{T5}, but does not improve on the bound in 
table \ref{T3}.

\begin{table}[h!]
\centering
\caption{Bounds on the comb growth constant  $\exp{\zeta(m_a,m_b)}$ by theorem \BRef{thm10}}
\lineup
\begin{tabular}{lccccccc}
\hline          
 & & Lattice & $m_a=1$ & $m_a=2$ & $m_a=3$ & $m_a=4$ & $m_a=5$ \cr 
\hline
\vspace{-2mm}
& & & & & \cr
\multirow{2}{*}{$m_b=1$} 
& & Square & $ 2.3849^*$ & $ 2.4206^*$ & $ 2.4538^*$ & $ 2.4962^*$ & $ 2.5170^*$\cr
& & Cubic & $ 4.4232^\dagger$ & $ 4.4918^\dagger$ & $ 4.5250^\dagger$ & $ 4.5559^\dagger$ & $ 4.5745^\dagger$ \cr \cr
\multirow{2}{*}{$m_b=2$} 
& & Square & $ 2.4371^*$ & $ 2.4715^*$ & $ 2.4695^*$ & $ 2.5014^*$ & $ 2.5139^*$ \cr
& & Cubic & $ 4.4968^\dagger$ & $ 4.5353^\dagger$ & $ 4.5431^\dagger$ & $ 4.5673^\dagger$ & $ $  \cr \cr
\multirow{2}{*}{$m_b=3$} 
& & Square & $ 2.5227^*$ & $ 2.5107^*$ & $ 2.6023^*$ & $ 2.5199^*$ & $ 2.5278^*$ \cr
& & Cubic & $ 4.5644^\dagger$ & $ 4.5676^\dagger$ & $ 4.7494$ & $ $ & $ $  \cr \cr
\multirow{2}{*}{$m_b=4$} 
& & Square & $ 2.5499^*$ & $ 2.5411^*$ & $ 2.5261^*$ & $ 2.6131^*$ & $ 2.5416^*$  \cr
& & Cubic & $ 4.5901^\dagger$ & $ 4.5916^\dagger$ & $ $ & $ $ & $ $  \cr \cr
\multirow{2}{*}{$m_b=5$} 
& & Square & $ 2.4794^*$ & $ 2.5591^*$ & $ 2.5441^*$ & $ 2.5509^*$ & $ $ \cr
& & Cubic & $ 4.6147^\dagger$ & $ $ & $ $ & $ $ & $ $  \cr \cr
\multirow{1}{*}{$m_b=6$} 
& & Square & $ 2.5893^*$ & $ 2.5004^*$ & $ 2.5569^*$ & $ $ & $ $ \cr
\hline
\multicolumn{7}{l}{$*$ -- less than lower bound \BRef{lower:mu},
and $\dagger$ -- less than best value \BRef{exact:mu}.} \\
\end{tabular}
\label{T6}
\end{table}

\section{Extension to brushes}
\label{sec:brushes}

In this section we look briefly at extensions of some of these ideas to brushes.  
A \emph{brush} is a modification of a comb where there are at least 2 teeth
at each branch point, so the backbone has $t$ vertices of degree $f$ where 
$4 \le f \le \gamma$.  All the teeth have $m_a$ edges.  We can obtain an upper bound 
on brushes by adaptations of the argument in section \ref{sec:short}.  
We write $b_f(m_a,m_b,t)$ for the number of brushes with $t$ vertices of degree $f$,
$t(f{-}2)$ teeth, all with $m_a$ edges, and $t{+}1$ backbone segments, 
all with $m_b$ edges. 

The arguments in section \ref{sec:long} can be adapted to work for brushes 
with only minor modifications.  By treating the teeth and backbone as independent 
self-avoiding walks, the upper bound in (\ref{eqn:maininequality}) becomes
\begin{equation}
b_f(m_a,m_b,t) \le c_{(t+1)m_b} {{\gamma{-}2}\choose{f{-}2}}^t 
\left(\frac{c_{m_a}}{\gamma}\right)^{(f-2)t}.
\end{equation}
The corresponding bound on the free energy is given by
\begin{equation}
\zeta_f(m_a,m_b) \le \frac{m_b \log \mu}{(f{-}2)m_a+m_b} 
+ \frac{\log {{\gamma{-}2}\choose{f{-}2}} 
- (f{-}2) \log \gamma + (f{-}2) \log c_{m_a}}{(f{-}2)m_a+m_b}.
\end{equation}

For the square lattice with $f=4$ the brush must have 2 teeth at each branch point.  
This gives 
\begin{equation}
\zeta_4(m_a,m_b) \le \frac{m_b \log \mu_2}{2 m_a+m_b} + \frac{2 \log (c_{m_a}/4)}{2 m_a+m_b}.
\end{equation}
Rearranging gives
\begin{equation}
\zeta_4(m_a,m_b) \le \log \mu_2 + \frac{2 \log (c_{m_a}/4) - 2 m_a \log \mu_2}{2m_a + m_b}.
\end{equation}
$\zeta_4(m_a,m_b) < \log \mu_2$ if $c_{m_a}/4 < 2.625622^{m_a} \le \mu_2^{m_a}$, 
using the bound on $\mu_2$ in equation \BRef{lower:mu} 
\cite{ConwayGuttmann}.   This is true for $m_a \leq 23$, for any  $m_b$ such that $1 \le m_b < \infty$..

\begin{theo}
In the square lattice, for $1\leq m_a \leq 23$ and 
$1\leq m_b < \infty$, $\zeta_4(m_a,m_b) < \kappa_2$. \hfill \qed
\end{theo}

For the simple cubic lattice $\gamma=6$ and $f=4,5$ or $6$. 
$\zeta_f(m_a,m_b) < \log \mu_3$ if
\begin{equation}
{{4}\choose{f{-}2}} (c_{m_a}/6)^{f-2} < 4.225^{(f-2)m_a} \le \mu_3^{(f-2)m_a}
\end{equation}
where we have used the bound on $\mu_3$ in \cite{FisherSykes}.  Using the
values of $c_n$ from \cite{SBB11} we find that this condition is satisfied if 
$m_a \le 4$ when $f=4$, when $m_a \le 7$ when $f=5$ and when $m_a \le 11$ 
when $f=6$, for all $m_b$ such that $1 \le m_b < \infty$.

\begin{theo}
In the cubic lattice, for $1\leq m_b < \infty$, the free energy $\zeta_f (m_a,m_b) < \kappa_3$ 
when $1\leq m_a\leq 4$ if $f=4$, $1\leq m_a\leq 7$ if $f=5$ and $1\leq m_b\leq 11$
if $f=6$. \hfill \qed
\end{theo}

\begin{figure}[h!]
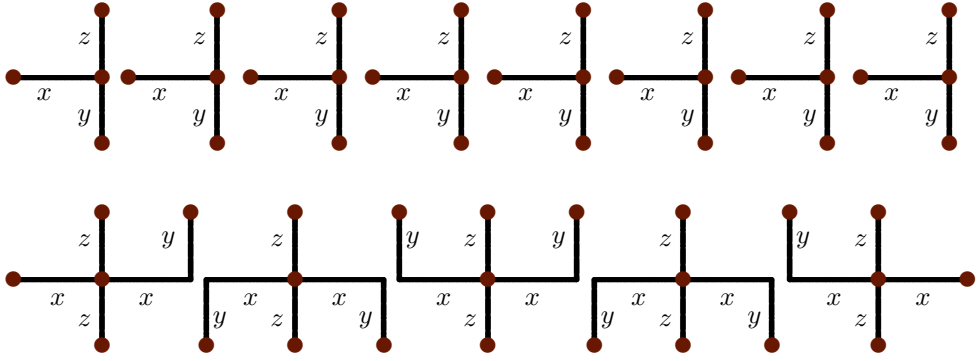

    \beginpicture
\setcoordinatesystem units <1.67pt,1.25pt>
\setplotarea x from -15 to 270, y from -30 to 30

\setplotsymbol ({\scalebox{0.50}{$\bullet$}})
\color{black}
\plot 0 0 20 0 /
\plot 20 -20 20 20 /
\put {$x$} at 7 -5   \put {$y$} at 16 -12   \put {$z$} at 16 12    
\color{Sepia}
\multiput {\scalebox{1.5}{$\bullet$}} at 0 0 20 0 20 -20 20 20   /

\multiput {
\beginpicture
\setplotsymbol ({\scalebox{0.50}{$\bullet$}})
\color{black}
\plot  0 0 20 0 /
\plot  20 -20 20 20 / 
\put {$x$} at 7 -5  \put {$y$} at 16 -12   \put {$z$} at 16 12    
\color{Sepia}
\multiput {\scalebox{1.5}{$\bullet$}} at 0 0 20 0 20 -20  20 20  / 
\endpicture
} at 35 0 62.5 0 90 0  117.5 0  145 0  172.5 0  200 0  /

\color{black}\normalcolor
\endpicture

    \beginpicture
\setcoordinatesystem units <1.67pt,1.25pt>
\setplotarea x from 5 to 270, y from -30 to 30

\setplotsymbol ({\scalebox{0.50}{$\bullet$}})
\color{black}
\plot 20 0 60 0 /
\plot 40 -20 40 20 /
\plot 60 0 60 20 /  
 \put {$y$} at 55 12 
 \put {$x$} at 30 -6    \put {$x$} at 50 -6    
 \put {$z$} at 36 12  \put {$z$} at 36 -12  
\color{Sepia}
\multiput {\scalebox{1.5}{$\bullet$}} at 20 0 40 0 40 20 60 20 40 -20  /

\multiput {
\beginpicture
\setplotsymbol ({\scalebox{0.50}{$\bullet$}})
\color{black}
\plot  0 0 40 0 /
\plot  20 -20 20 20 /
\plot  40 0 40 -20 /
\plot    0 0 0 -20 / 
\put {$y$} at 36 -12 \put {$y$} at 3 -12
\put {$x$} at 10 -6   \put {$x$} at 30 -6    
\put {$z$} at 16 12   \put {$z$} at 16 -12    
\color{Sepia}
\multiput {\scalebox{1.5}{$\bullet$}} at 0 -20 40 -20 20 0 20 20 20 -20   / 
\endpicture
} at 82.5 0  170 0 /

\multiput {
\beginpicture
\setplotsymbol ({\scalebox{0.50}{$\bullet$}})
\color{black}
\plot  0 0 40 0 /
\plot  20 -20 20 20 /
\plot  40 0 40 20 /
\plot  0 0 0 20 / 
\put {$y$} at 36 12  \put {$y$} at 3 12 
\put {$x$} at 10 -6   \put {$x$} at 30 -6    
\put {$z$} at 16 12  \put {$z$} at 16 -12    
\color{Sepia}
\multiput {\scalebox{1.5}{$\bullet$}} at 0 20 40 20 20 0 20 20 20 -20   / 
\endpicture
} at 126 0 /

\setplotsymbol ({\scalebox{0.50}{$\bullet$}})
\color{black}
\plot 195 20 195 0 235 0 /
\plot 215 20  215 -20 /
\put {$y$} at 198 12
\put {$x$} at 205 -6  \put {$x$} at 225 -6
\put {$z$} at 211 12  \put {$z$} at 211 -12 
\color{Sepia}
\multiput {\scalebox{1.5}{$\bullet$}} at 195 20 235 0 215 -20 215 0 215 20  /

\color{black}\normalcolor
\endpicture
\caption{Constructing a lattice brush by concatenating $3$-stars or $4$-stars.}
\label{F7}
\end{figure}


We can also construct upper bounds on brushes by concatenating stars.  
In figure \ref{F7} we show two decompositions of brushes into lattice stars, amongst
many other ways.  The top sequence shows a $4$-brush decomposed into $3$-stars,
and the bottom a decomposition into $4$-stars.  In the square lattice the top 
decomposition (with $m_b=x$ and $m_a=y=z$) gives
\begin{equation}
\zeta_4(m_a,m_b) \leq ( \log (\varsigma([m_b,m_a,m_a])/4) + \log s([m_b,m_a,m_a]) )/(2\,m_a+m_b)
\label{eqn36}
\end{equation}
where $\varsigma([m_b,m_a,m_a])$ is again a symmetry factor which is equal
to $3$ if $m_a=m_b$ (and equal to $1$ otherwise) and the factor $(1/4)$ is the 
result of the fact that only one direction is available when concatenating a star 
to its predecessor.  

If the coordination number of the lattice is more than $4$ then 
using $f$-stars instead of $3$-stars in the top decomposition gives $(f{+}1)$-brushes.
In the $d$-dimensional cubic lattice, using $f$ stars with $3\leq f\leq 2d{-}1$ 
and the top decomposition in figure \ref{F7}, one similarly obtains the bound
\begin{equation}
\hspace{-2.5cm}
\zeta_{f{+}1}(m_a,m_b) \leq ( \log ((2d{-}f)\,\varsigma([m_b,m_a,\ldots,m_a])/2d )
+ \log s([m_b,m_a,\ldots,m_a]) )/((f{-}1)m_a+m_b) .
\end{equation}
Here, $\varsigma([m_b,m_a,\ldots,m_a]) = f$ if $m_a=m_b$, and it is equal to $1$
otherwise.  Other, similar inequalities, can be obtained for other lattices, using
the same kind of analysis.  Notice that this reduces to equation \BRef{eqn36}
if $d=2$. We state this as the following theorem.

\begin{theo}
If $3\leq f\leq 2d{-}1$ then the connective constant $\zeta_f(m_a,m_b)$ of 
$f{+}1$-brushes in the $d$-dimensional hypercubic lattice is bounded by
$$
\zeta_{f{+}1}(m_a,m_b) \leq 
( \log ((2d{-}f)\,\varsigma([m_b,m_a,\ldots,m_a])/2d)
+ \log s([m_b,m_a,\ldots,m_a]) )
/((f{-}1)\,m_a+\,m_b) ,
$$
where $\varsigma([m_a{+}m_b,m_a{+}m_b,m_a,m_a])=f$ if
$m_a=m_b$, and equal to $1$ otherwise. \hfill \qed
\label{thm13}
\end{theo}

In the square lattice with $m_a=m_b=1$ and $f=3$ one obtains the bound
$\zeta_4(1,1) \leq ( \log 3 )/3 < \log 1.4423$.  In the cubic lattice, one similarly
obtains $\zeta_4(1,1) < \log 3.1073$, $\zeta_5(1,1) < \log 2.1148$ and
$\zeta_6(1,1) < \log 1.3780$.  More estimates are shown in table \ref{A2},
using the data in tables \ref{3star2d}, \ref{4star2d}, \ref{3star3d} and
\ref{4star3d}.

\begin{table}[h!]
\centering
\caption{Upper bounds on the brush growth constant $\exp{\zeta_{f+1}(m_a,m_b)}$ by theorem \ref{thm13}}
\lineup
\begin{tabular}{lllcccccccc}
\hline          
 $f{+}1$ & & Lattice & & $m_b$ & $m_a=1$ & $m_a=2$ & $m_a=3$ & $m_a=4$ & $m_a=5$ & $m_a=6$ \cr 
\hline
\vspace{-2mm}
& & & & & \cr
\multirow{13}{*}{$4$} 
 & & \multirow{6}{*}{Square}
  & & 1 & $ 1.4423^*$ & $ 1.9037^*$ & $ 2.0703^*$ & $ 2.2207^*$ & $ 2.2929^*$ & $ 2.3596^*$ \cr
 & &   & & 2 & $ 1.7321^*$ & $ 1.9948^*$ & $ 2.1309^*$ & $ 2.2392^*$ & $ 2.3079^*$ & $ 2.3599^*$ \cr
 & &   & & 3 & $ 1.8722^*$ & $ 2.0775^*$ & $ 2.1627^*$ & $ 2.2618^*$ & $ 2.3126^*$ & $ 2.3633^*$ \cr
 & &   & & 4 & $ 2.0154^*$ & $ 2.1491^*$ & $ 2.2192^*$ & $ 2.2930^*$ & $ 2.3393^*$ & $ 2.3791^*$ \cr
 & &   & & 5 & $ 2.0982^*$ & $ 2.2080^*$ & $ 2.2522^*$ & $ 2.3192^*$ & $ 2.3542^*$ & $ 2.3917^*$ \cr
 & &   & & 6 & $ 2.1773^*$ & $ 2.2537^*$ & $ 2.2905^*$ & $ 2.3498^*$ & $ 2.3748^*$ & $ 2.4053^*$ \cr \cr
 & & \multirow{6}{*}{Cubic}
 & & 1 & $ 3.1073^*$ & $ 3.7342^*$ & $ 3.39793^*$ & $ 4.1525^*$ & $ 4.2461^\dagger$ & $ 4.3232^\dagger$ \cr
 & &   & & 2 & $ 3.5000^*$ & $ 3.8759^*$ & $ 4.0648^*$ & $ 4.1919^*$ & $ 4.2745^\dagger$ & $ 4.3370^\dagger$ \cr
 & &   & & 3 & $ 3.7091^*$ & $ 3.9874^*$ & $ 4.1183^*$ & $ 4.2280^\dagger$ & $ 4.2932^\dagger$ & $ 4.3506^\dagger$ \cr
 & &   & & 4 & $ 3.8850^*$ & $ 4.0750^*$ & $ 4.1801^*$ & $ 4.2652^\dagger$ & $ 4.3222^\dagger$ & $ 4.3697^\dagger$ \cr
 & &   & & 5 & $ 3.9941^*$ & $ 4.1455^*$ & $ 4.2226^*$ & $ 4.2967^\dagger$ & $ 4.3428^\dagger$ & $ 4.3860^\dagger$ \cr
 & &   & & 6 & $ 4.0900^*$ & $ 4.2021^*$ & $ 4.2658^\dagger$ & $ 4.3255^\dagger$ & $ 4.3662^\dagger$ & $ 4.4027^\dagger$ \cr 
\hline
\vspace{-2mm}
& & & & & \cr
\multirow{5}{*}{$5$} 
 & & \multirow{5}{*}{Cubic}
 & & 1 & $ 2.1148^*$ & $ 3.0142^*$ & $ 3.3952^*$ & $ 3.6736^*$ & $ 3.8326^*$ & $ 4.2633^\dagger$ \cr
 & &  & & 2 & $ 2.5119^*$ & $ 3.1678^*$ & $ 3.4866^*$ & $ 3.7168^*$ & $ 3.8628^*$ & \cr
 & &  & & 3 & $ 2.7703^*$ & $ 3.3001^*$ & $ 3.5515^*$ & $ 3.7587^*$ & $ 3.8848^*$ &  \cr
 & &  & & 4 & $ 3.0080^*$ & $ 3.4193^*$ & $ 3.6322^*$ & $ 3.8081^*$ &  & \cr
 & &  & & 5 & $ 3.1765^*$ & $ 3.5203^*$ & $ 3.6933^*$ & $ 3.8515^*$ &  & \cr
\hline
\vspace{-2mm}
& & & & & \cr
\multirow{5}{*}{$6$} 
 & & \multirow{5}{*}{Cubic}
 & & 1 & $ 1.3800^*$ & $ 2.3891^*$ & $ 2.8613^*$ & $ 3.2224^*$ & & \cr
 & &  & & 2 & $ 1.7100^*$ & $ 2.5331^*$ & $ 2.9482^*$ & $ 3.2649^*$ & & \cr
 & &  & & 3 & $ 1.9547^*$ & $ 2.6645^*$ & $ 3.0150^*$ &  &  &  \cr
 & &  & & 4 & $ 2.1961^*$ & $ 2.7925^*$ & $ 3.1016^*$ &  &  &  \cr
 & &  & & 5 & $ 2.3838^*$ & $ 2.9053^*$ & $ 3.1708^*$ &  &  &  \cr
\hline
\multicolumn{9}{l}{$*$ -- below lower bound \BRef{lower:mu},
and $\dagger$ -- less than best value \BRef{exact:mu}.} \\
\end{tabular}
\label{T8}
\end{table}

A similar analysis using the bottom decomposition in figure \ref{F7} gives, 
with $m_a=y=z$ and $m_b=x$, the upper bound
\begin{equation}
\hspace{-2.0cm}
\zeta_4(m_a,m_b) \leq 
( \log (2\,(d{-}1)/d)
+ \log s([m_a{+}m_b,m_a{+}m_b,m_a,m_a]) )
/(8\,m_a{+}4\,m_b) .
\label{38}
\end{equation}
Recall that an extra node is created by each step in the concatenation.
Here, the factor of $(d{-}1)/d$ is due to the fact that $(2d{-}2)$ directions,
out of $2d$, are available for the edge on the part of the arm between the
node and the vertex where the star is concatenated. The symmetry factor 
$\varsigma([m_a{+}m_b,m_a{+}m_b,m_a,m_a]) = 2$ by default, for all 
values of $m_a$ and $m_b$.  Using this and the data in table \ref{nnmmstar2d}
gives $\zeta_4(1,1) \leq \log 1.3855$ in the square lattice, and the data
in table \ref{nnmmstar3d} gives $\zeta_4(1,1) \leq \log 1.9435$ in the
cubic lattice, improving on the bounds in table \ref{T8}.  Similarly, in
the square lattice this gives $\zeta_4(2,2) \leq \log 1.4790$, and in the cubic
lattice, $\zeta_4(2,2) \leq \log 2.0424$, as compared to the results in 
table \ref{T8}.

More decompositions may be done, but it remains unclear which would give
optimal bounds.  In general, the larger the basic component used to determine
the bound, and the extent to which intersections between the components
are avoided, the better the bound.

\section{Some limiting cases and lower bounds}
\label{sec:limiting}

In this section we shall examine situations where $m_a \to \infty$ 
or $m_b \to \infty$ \emph{after} the limit $t \to \infty$ has been taken.  

\subsection{Lower bounds}

To obtain lower bounds on $\zeta$ we shall consider the subset of combs on the 
simple cubic lattice, confined to a half-space ($z \ge 0$) and terminally attached 
to the dividing plane ($z=0$).  This subset of combs has been investigated in 
reference \cite{RensburgSoterosWhittington}.  

We first look at the case where $m_b \to \infty$ and $m_a \to \infty$ but with 
$m_a = o(m_b)$ so that the teeth grow less rapidly than the backbone segments.
This was called the \emph{self-avoiding walk limit} in \cite{RensburgSoterosWhittington}.
We can use the result of theorem 6 in \cite{RensburgSoterosWhittington} 
with $a=b=y=1$ to show that
\begin{equation}
\lim_{m_a,m_b \to \infty \atop{ m_a=o(m_b)}} \zeta(m_a,m_b) \ge \log \mu_3.
\end{equation}
The pattern theorem arguments used in section \ref{sec:long} do not apply 
when $m_b \to \infty$ since the pattern $P$ must be finite.

If $m_a \to \infty$ and $m_b \to \infty$ but with $m_b = o(m_a)$ so that 
the teeth grow more rapidly than the backbone segments,
this was called the \emph{star limit} in \cite{RensburgSoterosWhittington}.
We can use the result of theorem 4 in \cite{RensburgSoterosWhittington} 
with $a=b=y=1$ to show that
\begin{equation}
\lim_{m_a,m_b \to \infty \atop{ m_b=o(m_a)}} \zeta(m_a,m_b) \ge \log \mu_3.
\end{equation}

Next consider the case where $m_b = \lfloor \alpha p \rfloor$ and 
$m_a = \lfloor (1-\alpha)p \rfloor$ for some $\alpha \in (0,1)$.  If we let
$p \to \infty$ then both $m_a,m_b$ go to infinity.  Using the result of 
theorem 5 in \cite{RensburgSoterosWhittington} with $a=b=y=1$
(and remembering that $\kappa(1)=\lambda(1) = \log \mu_3$) we obtain the bound
\begin{equation}
\lim_{p\to\infty} \zeta(\lfloor (1-\alpha)p \rfloor, \lfloor \alpha p \rfloor) \ge \log \mu_3,
\end{equation}
for all $\alpha \in (0,1)$.

\subsection{Upper bounds}

We can obtain an upper bound by regarding the backbone as independent 
fom the teeth, and the teeth as independent of one another.  This gives
\begin{equation}
g(m_a,m_b,t) \le c_{(t+1)m_b} c_{m_a}^t
\end{equation}
and therefore
\begin{equation}
\fl \quad \quad
\frac{\log g(m_a,m_b,t)}{(m_a+m_b)t} 
\le \frac{m_b(t+1)}{(m_a+m_b)t} \; \frac{1}{m_b(t+1)} \log c_{(t+1)m_b} 
+\frac{m_a}{(m_a+m_b)} \frac{1}{m_a} \log c_{m_a}.
\end{equation}
Letting $t \to \infty$ gives
\begin{equation}
\zeta(m_a,m_b) 
\le \frac{m_b}{m_a+m_b} \log \mu_3 
  + \frac{m_a}{m_a+m_b} \; \frac{1}{m_a} \log c_{m_a}.
\end{equation}
We now consider three cases.  First, let $m_b \to \infty$ with $m_a=o(m_b)$.  
The first term goes to $\log \mu_3$ and the second term goes to zero.  Next, 
let $m_a \to \infty$ and $m_b = o(m_a)$.  In this case the first term 
goes to zero and the second term goes to $\log \mu_3$.  Finally, let 
$m_b = \lfloor \alpha p \rfloor$ and $m_a= \lfloor (1-\alpha)p \rfloor$ with 
$\alpha \in (0,1)$.  Then
\begin{equation}
\zeta(\lfloor (1-\alpha) p \rfloor,  \lfloor \alpha p \rfloor ) 
\le  \alpha  \log \mu_3 + (1-\alpha)  \; 
\frac{1}{\lfloor (1-\alpha)p \rfloor} \log c_{\lfloor (1-\alpha)p \rfloor}
\end{equation}
Letting $p \to \infty$ gives
\begin{equation}
\lim_{p\to\infty} \zeta(\lfloor (1-\alpha) p \rfloor, \lfloor \alpha p \rfloor) \le \log \mu_3.
\end{equation}

This shows that, for the three limiting cases that we have considered, 
$\zeta (m_a,m_b) = \log \mu_3$.  Similar arguments can be constructed for other lattices.

\section{Discussion}
\label{sec:discussion}

We have investigated the exponential growth rate of the numbers of combs embeddable in a lattice.  A comb consists of a backbone self-avoiding walk, and a number of teeth attached to the backbone at vertices of degree 3.  The comb has $t$ teeth, all with $m_a$
edges, with $m_b$ edges between adjacent vertices of degree 3 and between the end vertices of degree 3 and the end vertices of degree 1 of the backbone.
We have been mainly interested in the case where $m_a$ and $m_b$ are fixed and $t$ goes to infinity, and we have derived upper bounds on the exponential growth rate, using several different approaches, that show that the growth rate is strictly less than that of self-avoiding walks, when $m_a$ is small.

To derive upper bounds on the exponential growth rate, we have used two main approaches.  The first is to allow the backbone and teeth to intersect, and to
allow the teeth to intersect one another.  This is an extension of an approach due to Lipson \cite{Lipson1993}.
The second approach that we have used is to break up the comb into small lattice clusters (such as stars or walks) which are not necessarily mutually avoiding.  The two approaches give different upper bounds and both are useful for various values of $m_a$ and $m_b$. In section
\ref{sec:brushes} we have extended these ideas to brushes, where more than one tooth is attached to the backbone at each branch point.

There are a number of contributing factors.  Our use of a pattern theorem \cite{Kesten} in section \ref{sec:long} shows that the teeth reduce the contribution to the entropy from the conformational freedom of the backbone.  The teeth may themselves contribute more entropy than they would as components of an infinite walk, since $c_n 
> \mu^n$ for small values of $n$.  This is offset by two factors.  The first is that the teeth must be mutually avoiding, and must avoid the backbone, resulting in a lower entropy
contribution.  The second is that the number of possibilities for the first step of each tooth is reduced by the fact that the backbone uses two of the possible positions at each degree 3 vertex.  These competing factors lead to a reduction in entropy when $m_a$ is small.  Whether this is true for all \emph{finite} values of $m_a$ is an open question.

\section*{Acknowledgement}
EJJvR acknowledges financial support from the Natural Sciences and Engineering 
Research Council of Canada (NSERC) [funding reference number: 
RGPIN-2020-06339].

\section*{References}
\bibliographystyle{plain}
\bibliography{combs.bib}

\vfill\eject

\renewcommand{\arraystretch}{1.33}
\begin{table}[h!]
\centering
\caption{Square and cubic lattice counts for use with theorem \ref{thm9}}
\lineup
\resizebox{\columnwidth}{!}{%
\begin{tabular}{c@{}*{8}{r}ccccc}
\hline          
  Lattice & & $m_b$ & $s([1,m_b,1{+}m_b])$ & $s([2,m_b,2{+}m_b])$ & $s([3,m_b,3{+}m_b])$ 
                    & $s([4,m_b,4{+}m_b])$ & $s([5,m_b,5{+}m_b])$ & $s([6,m_b,6{+}m_b])$ \cr
\hline
\vspace{-2mm}
& & & & & \cr
\multirow{6}{*}{Square} 
& & 1 & $36$ & $520$ & $3736$ & $28008$ & $204416$ & $10844680$ \cr
& & 2 & $520$ & $1820$ & $25152$ & $184632$ & $1327992$ & $9611952$ \cr
& & 3 & $3736$ & $25152$ & $83404$ & $1210472$ & $8499040$ & $61232600$ \cr
& & 4 & $28008$ & $184632$ & $1210472$ & $4323540$ & $60403784$ & $429869160$ \cr
& & 5 & $204416$ & $1327992$ & $8499040$ & $60403784$ & $207611972$ & $2947819568$ \cr
& & 6 & $10844680$ & $9611952$ & $61232600$ & $429869160$ & $2947819568$ & $10365217796$ \cr
\hline
\vspace{-2mm}
& & & & & \cr
\multirow{6}{*}{Cubic} 
& & 1 & $300$ & $13656$ & $305880$ & $6910968$ & $154683552$ & $3472414632$  \cr
& & 2 & $13656$ & $152052$ & $6691680$ & $149877672$ & $3328747416$ & $74339457504$  \cr
& & 3 & $305880$ & $6691680$ & $72597636$ & $3230794920$ & $71202415152$ & $1584611766696$  \cr
& & 4 & $6910968$ & $149877672$ & $3230794920$ & $35747005596$ & $1569889038456$ & $  $  \cr
& & 5 & $154683552$ & $3328747416$ & $71202415152$ & $1569889038456$ & $ $ & $ $  \cr
& & 6 & $3472414632$ & $74339457504$ & $1584611766696$ &  &  &   \cr
\hline
\end{tabular}
}
\label{A1}
\end{table}

\begin{table}[h!]
\centering
\caption{Square and cubic lattice counts for use with theorem \ref{thm10}}
\lineup
\resizebox{\columnwidth}{!}{%
\begin{tabular}{c@{}*{8}{r}ccccc}
\hline          
Lattice & & $m_b$ & $s([m_b,1,[m_b],1{+}m_b,1])$ & $s([m_b,2,[m_b],2{+}m_b,2])$ & $s([m_b,3,[m_b],3{+}m_b,3])$ 
  & $s([m_b,4,[m_b],4{+}m_b,4])$ & $s([m_b,5,[m_b],5{+}m_b,5])$ \cr
\hline
\vspace{-2mm}
& & & & & \cr
\multirow{6}{*}{Square} 
& & 1 & $184$ & $5704$ & $95288$ & $1820424$ & $32866608$ \cr
& & 2 & $6064$ & $51928$ & $1548488$ & $29395672$ & $510888032$ \cr
& & 3 & $132808$ & $1985048$ & $29940240$ & $536787264$ & $9261420152$ \cr
& & 4 & $2504896$ & $39022296$ & $565536024$ & $10273542256$ & $173177851984$ \cr
& & 5 & $25064528$ & $742310824$ & $10804335600$ & $191214770848$ \cr
& & 6 & $950270624$ & $7129667536$ & $203746272232$ & $ $ \cr
\hline
\vspace{-2mm}
& & & & & \cr
\multirow{5}{*}{Cubic} 
& & 1 & $5616$ & $1116360$ & $110533032$ & $11341084056$ & $1154028733584$  \cr
& & 2 & $1127568$ & $56787648$ & $10870851288$ & $1121853763176$   \cr
& & 3 & $122649960$ & $11786176440$ & $1133744588640$ & $ $   \cr
& & 4 & $12688738992$ & $1233919271304$ & $ $ & $ $   \cr
& & 5 & $1350810161904$ & $ $ & $ $ & $ $   \cr
\hline
\end{tabular}
}
\label{A2}
\end{table}

\renewcommand{\arraystretch}{1.33}
\begin{table}[h!]
\centering
\caption{Square lattice 3-star counts }
\lineup
\begin{tabular}{l@{}*{12}{r}}
\hline          
 \scalebox{0.9}{$n$}  &  &
 \scalebox{0.9}{$s([n,1,1])$} & 
 \scalebox{0.9}{$s([n,2,2])$} & 
 \scalebox{0.9}{$s([n,3,3])$} & 
 \scalebox{0.9}{$s([n,4,4])$} & 
 \scalebox{0.9}{$s([n,5,5])$} & 
 \scalebox{0.9}{$s([n,6,6])$}  & \cr
\hline
1  & & 4 & 100 & 652 & 5252  & 36836 & 281132 \cr
2  & & 36 & 84 & 1700 & 12676 & 91316 & 664324  \cr
3  & & 92 & 668 & 1380 & 31692 & 216452 & 1602428 \cr
4  & & 268 & 1820 & 11588 & 28164 & 587796  & 4211372 \cr
5  & & 716 & 4988 & 30244 & 224620 & 504084 & 10961516 \cr
6  & & 2020 & 13516 & 83404 & 600892 & 4093036 & 9675548 \cr
\hline
\end{tabular}
\label{3star2d}
\end{table}

\begin{table}[h!]
\centering
\caption{Square lattice 4-star counts }
\lineup
\begin{tabular}{l@{}*{12}{r}}
\hline          
 \scalebox{0.9}{$n$}  &  &
 \scalebox{0.9}{$s([n,1,1,1])$} & 
 \scalebox{0.9}{$s([n,2,2,2])$} & 
 \scalebox{0.9}{$s([n,3,3,3])$} & 
 \scalebox{0.9}{$s([n,4,4,4])$} & 
 \scalebox{0.9}{$s([n,5,5,5])$} & 
 \scalebox{0.9}{$s([n,6,6,6])$}  & \cr
\hline
1  & & 1 & 84 & 1036 & 23716 & 396052 & 8064684  \cr
2  & & 12 & 47 & 2420 & 48716 & 849228 & 16110364 \cr
3  & & 28 & 460 & 1297 & 107548 & 1712684 & 33264108 \cr
4  & & 84 & 1252 & 14588 & 70257 & 4594732 & 84483172 \cr
5  & & 220 & 3404 & 36676 & 725732 & 2809521 & 210089420 \cr
6  & & 628 & 9204 & 101556 & 1915436 & 30168356 & 136065237 \cr
\hline
\end{tabular}
\label{4star2d}
\end{table}

\renewcommand{\arraystretch}{1.33}
\begin{table}[h!] 
\centering
\caption{Cubic lattice 3-star counts }
\lineup
\begin{tabular}{l@{}*{12}{r}}
\hline          
 \scalebox{0.9}{$n$}  &  &
 \scalebox{0.9}{$s([n,1,1])$} & 
 \scalebox{0.9}{$s([n,2,2])$} & 
 \scalebox{0.9}{$s([n,3,3])$} & 
 \scalebox{0.9}{$s([n,4,4])$} & 
 \scalebox{0.9}{$s([n,5,5])$} & 
 \scalebox{0.9}{$s([n,6,6])$}  & \cr
\hline
1  & & 20 & 1452 & 31596 & 734196  & 16174908 & 368476164  \cr
2  & & 300 & 2260 & 149052 & 3350604 & 74400420  & 1665999876   \cr
3  & & 1404 & 32052 & 227172 & 15434316 & 336611868 & 7572405156  \cr
4  & & 6876 & 152052 & 3257412 & 24159620 & 1588051596 & 35328582708  \cr
5  & & 32436 & 722988 & 15219372  & 340216500  & 2456804660 & 164468765436  \cr
6  & & 156588 & 3432684 & 72597636  & 1605227796 & 34882592748 & 257495456028  \cr
\hline
\end{tabular}
\label{3star3d}
\end{table}

\renewcommand{\arraystretch}{1.33}
\begin{table}[h!]
\centering
\caption{Cubic lattice 4-star counts }
\lineup
\begin{tabular}{l@{}*{12}{r}}
\hline          
 \scalebox{0.9}{$n$}  &  &
 \scalebox{0.9}{$s([n,1,1,1])$} & 
 \scalebox{0.9}{$s([n,2,2,2])$} & 
 \scalebox{0.9}{$s([n,3,3,3])$} & 
 \scalebox{0.9}{$s([n,4,4,4])$} & 
 \scalebox{0.9}{$s([n,5,5,5])$} & 
 \scalebox{0.9}{$s([n,6,6,6])$}  & \cr
\hline
1  & & 15 & 6780 & 610620 & 66573996 & 6500046708 & 691980258780  \cr
2  & & 300 & 7605 & 2776140 & 287996100 & 28469950428 &  \cr
3  & & 1356 & 139260 & 3019767 & 1266316836  & 121806730620 &  \cr
4  & & 6684 & 655332 & 57451812 & 1466438355  & & \cr
5  & & 31092 & 3085884 & 263549340 & 27088915788 & & \cr
\hline
\end{tabular}
\label{4star3d}
\end{table}

\renewcommand{\arraystretch}{1.33}
\begin{table}[h!]
\centering
\caption{Square lattice 4-star counts }
\lineup
\begin{tabular}{l@{}*{12}{r}}
\hline          
 \scalebox{0.9}{$n$}  &  &
 \scalebox{0.9}{$s([n,n,1,1])$} & 
 \scalebox{0.9}{$s([n,n,2,2])$} & 
 \scalebox{0.9}{$s([n,n,3,3])$} & 
 \scalebox{0.9}{$s([n,n,4,4])$} & 
 \scalebox{0.9}{$s([n,n,5,5])$} & 
 \scalebox{0.9}{$s([n,n,6,6])$}  & \cr
\hline
1  & & 1 & 50 & 270 & 2322 & 15594 & 122522  \cr
2  & &  & 47 & 1618 & 11998 & 84942 & 613958 \cr
3  & &  &  & 1297 & 60146  & 380610 & 2839642 \cr
4  & &  &  &  & 70257 & 2764030 & 19260478 \cr
5  & &  &  &  &  & 2809521 & 120128342 \cr
6  & &  &  &  &  &  & 136065237 \cr
\hline
\end{tabular}
\label{nnmmstar2d}
\end{table}

\renewcommand{\arraystretch}{1.33}
\begin{table}[h!]
\centering
\caption{Cubic lattice 4-star counts }
\lineup
\begin{tabular}{l@{}*{12}{r}}
\hline          
 \scalebox{0.9}{$n$}  &  &
 \scalebox{0.9}{$s([n,n,1,1])$} & 
 \scalebox{0.9}{$s([n,n,2,2])$} & 
 \scalebox{0.9}{$s([n,n,3,3])$} & 
 \scalebox{0.9}{$s([n,n,4,4])$} & 
 \scalebox{0.9}{$s([n,n,5,5])$} & 
 \scalebox{0.9}{$s([n,n,6,6])$}  & \cr
\hline
1  & & 15 & 2178 & 44130 & 1040526 & 22297842 & 512569842  \cr
2  & &  & 7605 & 940602 & 20800758 & 453100410 & 10082409522 \cr
3  & &  &  & 3019767 & 406377102 & 8541846738 & 191746501074 \cr
4  & &  &  &  & 1466438355 & 186596132166  &  \cr
\hline
\end{tabular}
\label{nnmmstar3d}
\end{table}


\vspace{4cm}
\vfill\eject
\end{document}